\documentclass[12pt,preprint]{aastex}

\usepackage{emulateapj5}

\begin {document}

\title{The Role of Evolutionary Age and Metallicity in the Formation 
of Classical Be Circumstellar Disks II. Assessing the Evolutionary 
Nature of Candidate Disk Systems}

\author{J.P. Wisniewski\altaffilmark{1,2,3}, K.S. Bjorkman\altaffilmark{3,4}, 
A.M. Magalh\~aes\altaffilmark{3,5}, J.E. Bjorkman\altaffilmark{4}, 
M.R. Meade\altaffilmark{6}, \& Antonio Pereyra\altaffilmark{5}}

\altaffiltext{1}{NASA GSFC Exoplanets and Stellar Astrophysics Lab Code 667, 
Greenbelt, MD 20771 USA John.P.Wisniewski@nasa.gov}
\altaffiltext{2}{NPP Fellow}
\altaffiltext{3}{Visiting Astronomer, Cerro Tololo Inter-American Observatory}
\altaffiltext{4}{Department of Physics and Astronomy 
MS 113, University of Toledo, Toledo, OH 43606-3390 USA, Karen.Bjorkman@.utoledo.edu, Jon.Bjorkman@utoledo.edu}
\altaffiltext{5}{IAG, Universidade de S\~ao Paulo, Caixa Postal 3386, S\~ao Paulo, SP 
01060-970, Brazil, mario@astro.iag.usp.br, antonio@astro.iag.usp.br}
\altaffiltext{6}{Space Astronomy Lab, University of Wisconsin-Madison,
1150 University Avenue, Madison, WI 53706 USA, meade@sal.wisc.edu}

\begin{abstract}

We present the first detailed imaging polarization observations of six SMC and six LMC clusters, known 
to have large populations of B-type stars which exhibit excess H$\alpha$ emission, to 
constrain the evolutionary status of these stars and hence better establish links between the onset of disk 
formation in classical Be stars and cluster age and/or metallicity.  We parameterize the 
interstellar polarization (ISP) along the lines of sight to these twelve clusters, thereby providing a diagnostic
of the fundamental properties of the dust which characterizes their localized interstellar medium.  We determine that the ISP associated with the 
SMC cluster NGC 330 is characterized by a modified Serkowski law with $\lambda_{max}$ 
$\sim$4500\AA, indicating the presence of smaller than average dust grains.  Furthermore, the morphology of the 
ISP associated with the LMC cluster NGC 2100 suggests that its interstellar environment is characterized by 
a complex magnetic field.    

Removing this interstellar polarization component from our data isolates the presence of any intrinsic polarization; 
the wavelength dependence of this intrinsic polarization provides a diagnostic of the dominant and any secondary polarigenic agents present, enabling us to discriminate pure gas disk systems, i.e. classical Be stars, from composite gas plus dust disk systems, i.e. Herbig Ae/Be or B[e] stars.  Our \textit{intrinsic} polarization results, along with available near-IR color information, strongly supports
 the suggestion of Wisniewski et al. that classical Be stars are present in clusters of age 5-8 Myr, and 
contradict assertions that the Be phenomenon only develops in the second half of a B star's main sequence 
lifetime, i.e. no earlier than 10 Myr.  Our data imply that a significant number of B-type stars must 
emerge onto the zero-age-main-sequence rotating at near-critical rotation rates, although we can not rule out the 
possibility that these data instead reveal the presence of a sub-group of the Be phenomenon characterized by 
sub-critically rotating objects.  

Comparing the polarimetric properties of our dataset to a similar survey of Galactic classical Be stars, we find that 
the prevalence of polarimetric Balmer jump signatures decreases with metallicity.  We 
speculate that these results might indicate that either it is more difficult to form large disk systems in low metallicity 
environments, or that the average disk temperature is higher in these low metallicity environments.  We have 
characterized the polarimetric signatures of all candidate Be stars in our data sample and find $\sim$25\% 
are unlikely to arise from true classical Be star-disk systems.  This detection of such a substantial number 
``contaminants'' suggests one should proceed with caution when attempting to determine the role of evolutionary 
age and/or metallicity in the Be phenomenon purely via 2-CD results.

\end{abstract}


\section{Introduction}

While the rapid rotation (v$_{eq}$/v$_{crit}$ $\sim$70-80\% of their critical 
velocity; \citealt{por96, por03}) of classical Be stars has long been speculated to be 
the fundamental source driving the production of their geometrically thin 
circumstellar disks \citep{str31, por03}, recent photometric surveys \citep{fea72,gre92,gre97,
die98,kel99,gre00,kel00,ols01,mcs05,wi06a} have suggested secondary mechanisms might 
contribute to the observed phenomenon.  Specifically, \citet{mer82, gre97, fab00} and \citet{kel04} 
found that the frequency of the Be phenomenon seems to peak in clusters 
with a main sequence turn-off of B1-B2, leading to the suggestion that the Be phenomenon is
enhanced with evolutionary age.  Several recent observational studies 
\citep{gre92,maz96,gre97,mae99,kel04} have also suggested that
the Be phenomenon may be more prevalent in low metallicity environments, based on comparisons
of the fractional Be populations of Galactic, LMC, and SMC clusters.  

In \citet{wi06a} (hereafter
WB06), we used a simple 2-color diagram (2-CD) technique to identify the 
fractional candidate Be population of numerous Large Magellanic Cloud (LMC), Small 
Magellanic Cloud (SMC), and Galactic clusters in an effort to improve the statistical database which 
has been used to link classical Be disk formation with evolutionary age and/or metallicity.  
WB06 found evidence that the Be phenomenon develops much earlier 
than previously predicted by theory \citep{fab00}, i.e before the mid-point main sequence lifetime; furthermore, they found evidence of an additional enhancement in the fractional Be content of clusters 
with evolutionary age.  The increased statistics offered by this work, while confirming the previously
suggested trend of an enhancement in the Be phenomenon in low metallicity environments, lowered 
the average fractional Be content of SMC clusters from 39\% \citep{mae99} to 32\%.     

While the 2-CD has been widely used to link evolutionary age and/or metallicity with Be disk
formation, it is inherently unclear whether all B-type objects identified as excess H$\alpha$ 
emitters, i.e. ``Be stars'', are truly classical Be stars.  It has been noted that other
B-type objects, such as Herbig Ae/Be stars, post-main-sequence B[e] stars, and
supergiants, may also exhibit H$\alpha$ emission hence ``contaminate'' these claimed 
detections (WB06). 

Polarimetry is a tool which has long been used to investigate the circumstellar environments of
Be stars \citep{coy69, poe76, coy82, bjo94, mcd94, qui97, woo97, cla98}.  Electron scattering 
in the innermost region of classical Be circumstellar disks will polarize a small fraction of stellar
photons; furthermore, pre- and post-scattering attenuation by hydrogen atoms within the
inner disk region will imprint a characteristic wavelength dependence onto this polarization 
signature if the inner disk is sufficiently dense \citep{woo95}.  The wavelength 
dependence of intrinsic polarization originating from classical Be stars, 
characterized by geometrically thin, gaseous circumstellar disks, \citep{qui97, woo97, cla98} is
significantly different than that expected from the dustier circumstellar environments of Herbig 
Ae/Be stars, post-main-sequence B[e] stars, and supergiants \citep{vra79,mey02}.  Thus for 
non-pole-on disk 
geometries, polarimetry can be employed to investigate circumstellar environments and
discriminate the evolutionary nature of sources. 

In this paper, we use imaging polarimetry to investigate the true nature of 
candidate Be stars in six LMC and six SMC clusters which were initially identified via 
photometric 2-CDs.  In Section 2, we 
outline our observations and data reduction techniques.  In Section 3, we detail how
we identified and removed the interstellar polarization component associated with each of 
our lines of sight.  We develop a classification system to characterize the intrinsic polarization 
observed in our candidate Be star sample in Section 4 and also discuss the results of 
individual clusters.   We discuss the implications of these results in Section 5 and summarize
our work in Section 6.

\section{Observations and Data Reduction}

The imaging polarimetry data presented in this study were obtained at the Cerro Tololo 
Inter-American Observatory (CTIO)\footnote{The Cerro 
Tololo Inter-American Observatory is operated by 
the Association of Universities for Research in Astronomy, under contract with the National Science 
Foundation.} 1.5m.
The f/7.5 secondary configuration was used during our 2001 observing run,
which yielded a 15 x 15 arc-minute field of view and a
0.44 arc-second pixel$^{-1}$ scale, while the 
f/13.5 secondary configuration, which provided a 8 x 8 arc-minute field of view with a 0.24
arc-second pixel$^{-1}$ scale, was used during our 2002 observing run.  Data were 
recorded with CTIO's standard Cassegrain focus CCD (CFCCD), a 2048 x 2048 
multi-amplifier CCD.  The standard CFCCD configuration was modified by the addition of a 
rotatable half-wave plate followed by a fixed analyzer, the latter of which was placed
in the upper of the 1.5m's two filter wheels.  The polymer achromatic
wave-plate was manufactured by Meadowlark Optics and provided 
polarization modulation for each 90$^{\circ}$ rotation of the
wave-plate.  Unless otherwise noted, each of our fields of view was 
observed with the wave-plate rotated through 8 positions 22.5$^{\circ}$ 
apart.  The fixed analyzer was a double calcite block whose
optical axes had been crossed to minimize astigmatism and color effects.  
Further details about this polarimeter can be found in 
\citet{mag96}, \citet{per00}, \citet{mel01}, \& \citet{per02}.  We placed 
CTIO's Tek U, B, V, R, and I filters in the second filter wheel.
A summary of the science clusters we observed is presented in Table \ref{polexpose}.

Nightly observations of polarization standard stars were obtained to calibrate
the polarimetric efficiency of each filter and determine the zero-point of the polarization position
angle in each filter, while nightly observations of unpolarized standard stars were obtained 
to characterize any instrumental polarization present.  Our unpolarized standard data
indicate that the polarimeter had an instrumental polarization consistent with
zero for our 2001 data.  The B, V, R, and I filters of our 2002 data also had an instrumental 
polarization consistent with zero \citep{wis03}, while the U filter data had 
an instrumental polarization of 
$\sim$0.1\%, which was removed from the data.

The reduction of these data began with standard image processing, including bias
and flat field corrections, using standard 
IRAF\footnote{IRAF is distributed by the National Optical Astronomy 
Observatories, which are operated by the Association of Universities for 
Research in Astronomy, Inc., under contract with the National Science 
Foundation.} techniques.  Following this initial processing, aperture 
photometry was performed for all point sources in each of the 8 images of 
an observation set, using 10 apertures of size 3-12 pixels for 
the f/7.5 data and 5-14 pixels for the f/13.5 data.  The linear
polarization of each source was extracted from a least squares solution of the difference 
amplitudes in the 8 wave-plate positions ($\psi_{i}$), using the PCCDPACK
software suite \citep{per00,per02}.  Both the expected
photon noise errors and the actual measured errors, defined as the
residuals at each wave-plate position with respect to the expected
$\cos 4\psi_{i}$ curve, were also calculated.  With few exceptions, these 
two errors were consistent with one another.  

\section{Interstellar Polarization}

The polarization observed in our raw data is comprised of the 
superposition of an interstellar component, attributable to the dichroic absorption of starlight 
by partially aligned dust grains along each line of sight, and possibly an intrinsic component, attributable  
to the asymmetric illumination of circumstellar material and/or the illumination of an asymmetrical distribution 
of circumstellar material.  Characterizing and isolating each of these components, although often technically challenging, yields two unique diagnostics.  Isolating the interstellar polarization 
component can facilitate detailed studies of the local interstellar medium (ISM) 
and provide constraints on fundamental properties such as grain size distributions and shapes, as well 
as diagnosing the magnitude and direction of local 
magnetic fields.  Isolating the intrinsic polarization component enables investigations of the 
distribution and chemistry of scatterers which comprise unresolved circumstellar environments.  
We stress that characterizing the polarizing agent (i.e. gas or dust) responsible for producing an 
intrinsic polarization signal is not possible simply via inspection of the wavelength dependence of 
the vector sum of interstellar plus intrinsic (i.e. the total) polarization data.  

Fortunately, our efforts to identify and separate the interstellar and intrinsic polarization components 
of our data are simplified by the fact that we have observed rich stellar associations.  Most of the stars in 
these clusters should be normal main sequence stars lacking any type of asymmetrical circumstellar envelope, hence they should not exhibit an intrinsic polarization component. Furthermore, we can 
assume that all of our stars are located at the same general distance, since they are members of a 
cluster population, and that the properties 
of the interstellar medium do not significantly vary over the angular extent of our clusters.  Hence 
each cluster in our sample provides us with a multitude of suitable ``field stars'' whose polarization we
can simply average to estimate the interstellar polarization (ISP) along each line of sight, via the 
commonly used field star technique \citep{mcl78}.  

\begin{figure}
\begin{center}
\includegraphics[scale=1.0]{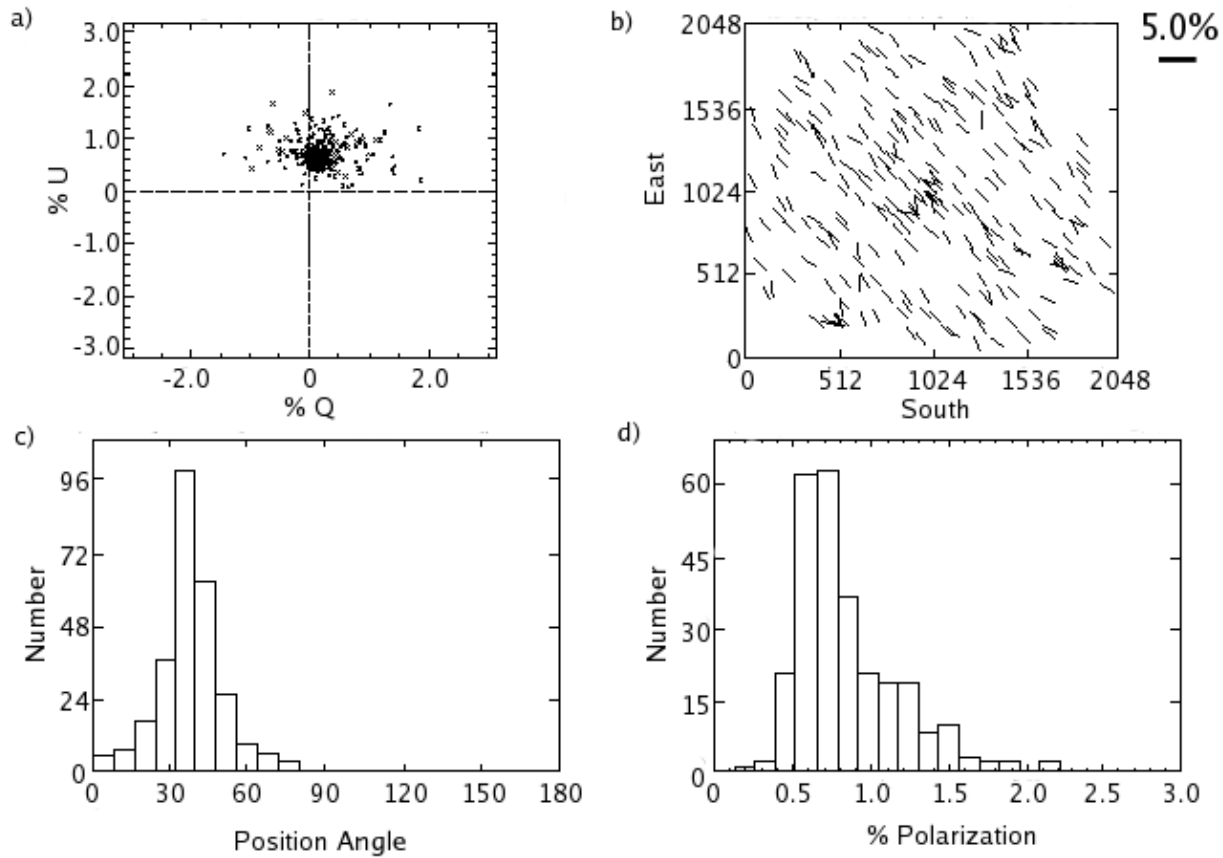}
\caption{The output of the 
PCCDPACK routine \textit{select} for the V-band observation of
NGC 1818 is shown.  Cluster members having a polarimetric 
signal-to-noise level, p/$\sigma_{p}$, greater than 5 are plotted on a Stokes Q-U
diagram (panel a), as a polarization map as a function of pixel location (panel b), and 
in position angle (panel c) and polarization (panel d) histograms.  As many cluster stars in 
panels a, c, and d exhibit the same general polarimetric properties, we suggest that most of these 
objects lack significant intrinsic polarization components and instead only exhibit evidence of  
interstellar polarization. \label{1818vsel}}
\end{center}
\end{figure}

\begin{figure}
\begin{center}
\includegraphics[scale=0.5]{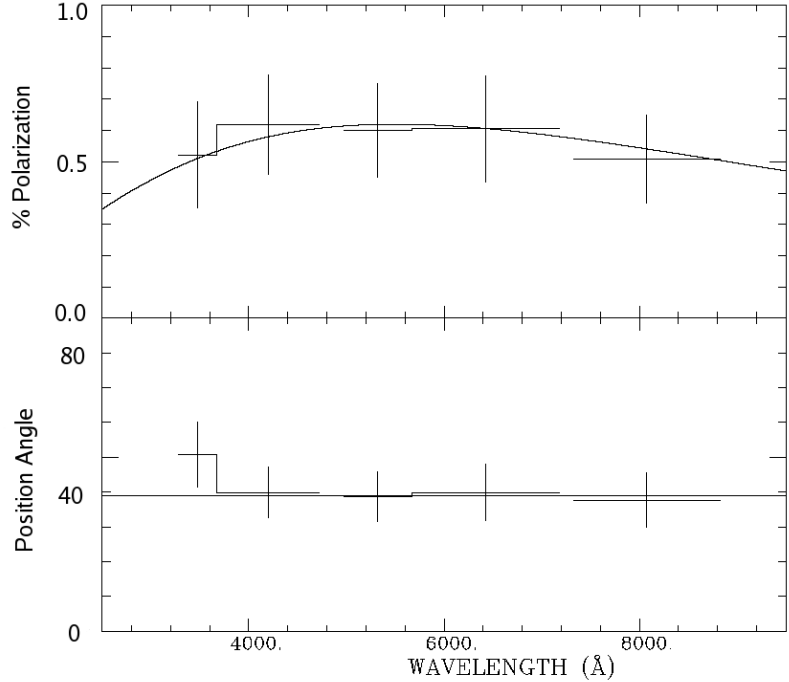}
\caption{The observationally derived total ISP estimates for 
each filter of NGC 1818 are plotted as
a function of wavelength.  The data closely follow a classic Serkowski-law
shape \citep{ser75}, suggesting that our method for determining the ISP in
each cluster was reasonable.  Over-plotted on these estimates is a 
modified Serkowski-law \citep{ser75,wil82}
which we deemed best fit the estimates.  We extracted the U, B, V, R, and I filter 
polarization associated with this fit, and hereafter use these values to 
describe the ISP of NGC 1818. \label{1818serkv}}
\end{center}
\end{figure}

In Figure \ref{1818vsel} we show the total V-band polarization of all sources in the LMC cluster NGC 1818 
having polarimetric signal-to-noise ratios (p/$\sigma_{p}$) greater than 5.0. 
The distinctive grouping of most objects in the Stokes Q-U diagram of this 
figure (panel a) and the narrow polarization (panel d) and position angle (panel c) histograms demonstrates 
that most of these objects do lack significant intrinsic polarization components and only have an ISP 
component.  After excluding objects outside of the dominant trends found in these histograms with an
iterative method, we calculated the weighted average and standard deviation of these data to determine
a preliminary estimate of this cluster's V-band ISP, as shown in Figure \ref{1818serkv}.  The 
wavelength dependence of these estimates clearly follow the 
empirical Serkowski law \citep{ser75} commonly used to parameterize an ISP.  Over-plotted in Figure 
\ref{1818serkv} is a modified Serkowski law \citep{ser75, wil82} which we deemed best represented the
observational data.  The polarization at the midpoint wavelength location
of the U, B, V, R, and I filters was then extracted from this curve and served as the final 
estimate of the ISP.  The modified Serkowski parameters determined for 
all of our lines of sight, along with the extracted filter ISP values, are summarized in Table \ref{isptable}.

The ISP values compiled in Table \ref{isptable} are relevant to our efforts to isolate the intrinsic 
polarization components associated with our target stars; however, they do not yet 
yield any direct information regarding 
the properties of the interstellar dust grains which reside in their parent LMC and SMC clusters, 
as these ISP values are comprised of both foreground 
Galactic ISP and Magellanic Cloud ISP contributions.  Following the techniques used by 
\citet{cla83} and \citet{rod97}, we used the interstellar polarization maps of \citet{sch76}, to identify and remove this 
foreground Galactic ISP contribution (see Table \ref{foreground}), hence determine estimates of the ISP 
intrinsic to our LMC and SMC clusters (see Table \ref{ispintrin}).  We used a modified Serkowski-law \citep{ser75,wil82} 
to extrapolate these maps over the wavelength range of our dataset, and assumed a nominal Galactic $\lambda_{max}$ 
value of 5500\AA\ to characterize the parameter K.   The errors cited in the Galactic polarization maps \citep{sch76} 
and those present in our total ISP estimates (Table \ref{isptable}) were propagated to produce the error estimates cited 
in Table \ref{ispintrin}.  

\begin{figure}
\begin{center}
\includegraphics[scale=0.5]{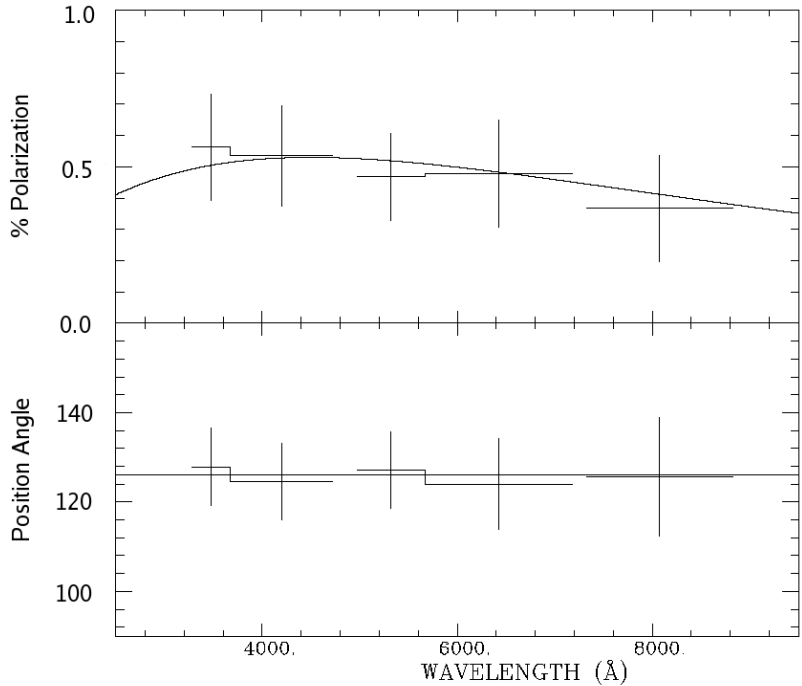}
\caption{The observationally derived total ISP estimates for 
each filter of NGC 330 are plotted as
a function of wavelength.  Over-plotted on these estimates is a 
modified Serkowski-law \citep{ser75,wil82}
which we deemed best fit the data.  These data, as well as the
SMC ISP associated with NGC 330 (see Table \ref{ispintrin}) are
characterized by a $\lambda_{max}$ value of 
$\sim$4500 \AA, suggesting the presence of smaller than average 
dust grains. \label{330serk}}
\end{center}
\end{figure}

Most of the field stars used to derive our total ISP estimates were not highly reddened objects, hence the intrinsic 
LMC and SMC ISP values listed in Table \ref{ispintrin} all are characterized by modest polarization amplitudes, which are
often on the order of the errors of our data.  In spite of this fact, trends in these data are clearly present.  
As expected, the magnitude of UBVRI polarization towards each cluster follows a Serkowski-like wavelength dependence
at a wavelength independent position angle.  This lack of position angle rotation indicates that our Galactic 
interstellar polarization correction was reasonable and that we are looking at a single magnetic field 
orientation towards each of our clusters.  All LMC clusters 
exhibit similar ISP properties, with polarization magnitudes ranging from $\sim$0.2-0.5\% at a position angle 
of $\sim$25-45 degrees; similarly, all SMC clusters also exhibit clear evidence of sharing common ISP 
properties, with polarization magnitudes ranging from $\sim$0.3-0.6\% at a position angle of 
$\sim$120-150 degrees.  The shallow curvature of the Serkowski-law dependence of these data, in combination with 
the moderate level of uncertainty present, make identifying systematic differences in the interstellar dust grain properties 
of these clusters difficult.  The total ISP (Table \ref{isptable}) and SMC ISP (Table \ref{ispintrin}) of the cluster 
NGC 330 does however show suggestive evidence of being characterized by 
a short wavelength $\lambda_{max}$ value of $\sim$4500 \AA\ (Figure \ref{330serk}).  Such an 
effect is commonly attributed to the 
presence of small dust grains, and has been previously observed in other SMC sight-lines \citep{rod97}.  In a future paper, we will examine the wavelength dependence of moderately and highly reddened objects in our dataset to perform a more detailed investigation of the interstellar medium properties of these 
LMC and SMC clusters, and compliment earlier studies which parameterized the ISM properties of other 
SMC \citep{rod97} and LMC \citep{cla83} sight-lines.

\subsection{NGC 2100}

\begin{figure}
\begin{center}
\includegraphics[scale=0.5]{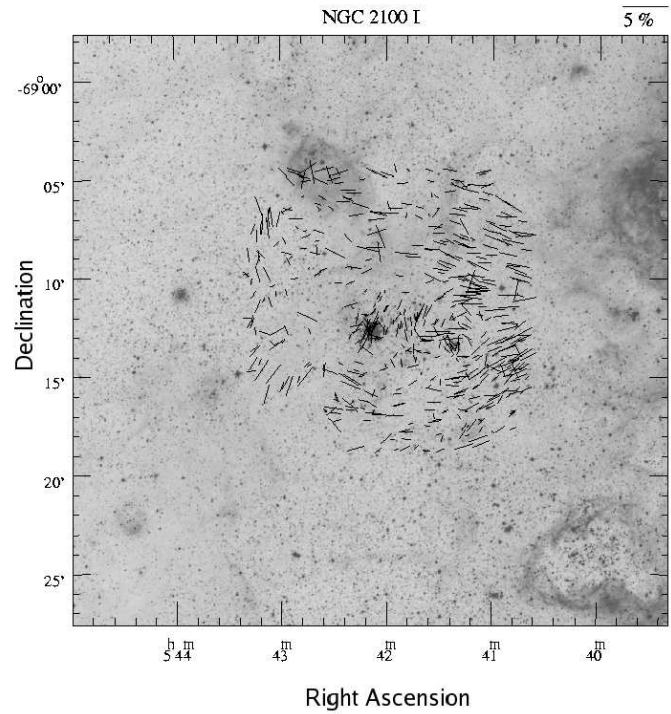}
\caption{The I-band total polarization vectors of our observations of
 the LMC cluster NGC 2100 are
over-plotted on a DSS-2 red image.  The general morphology of these polarization vectors matches
that seen in the other 4 filters observed, indicating this structure is real. \label{rawn2100i}}
\end{center}
\end{figure}

The total polarization vectors of our observations of the LMC cluster 
NGC 2100 clearly exhibit evidence of a collective, complex morphology (see Figure \ref{rawn2100i}).  
While Figure \ref{rawn2100i} only presents I-band data, all of our other filters exhibit similar 
alignment patterns.  Recall that we expect most cluster objects should not exhibit an 
intrinsic polarization component; thus 
we suggest that these systematic morphological changes are related to changes in
the magnetic field properties within our field of view.  A substantial discussion of the magnetic
field properties of NGC 2100 is presented in \citet{wi06b}.  For the purposes of the present study, we 
will only discuss our efforts to parameterize and remove the ISP along the line of sight to this cluster.

\begin{figure}
\begin{center}
\includegraphics[scale=0.5]{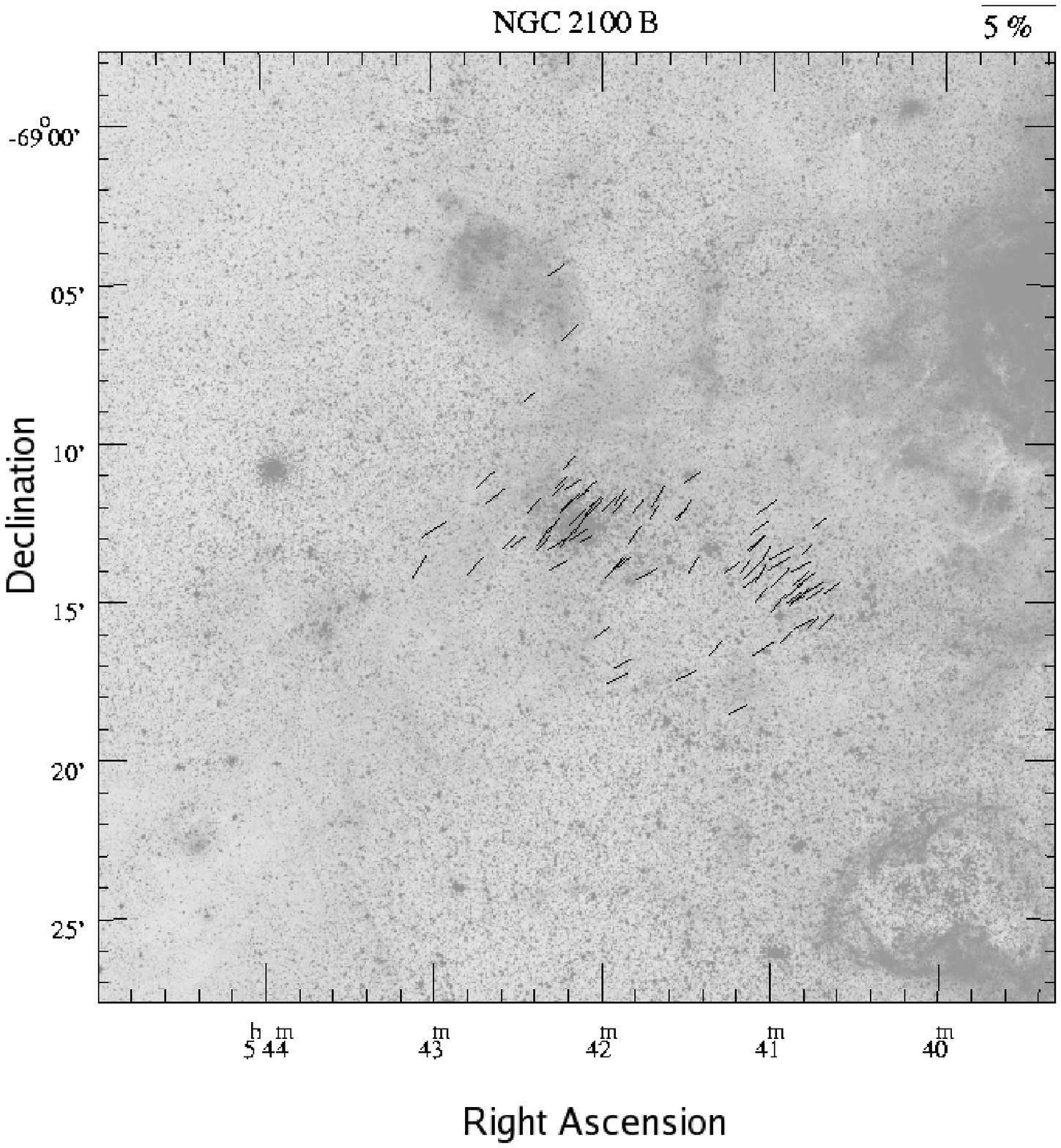}
\caption{The B-band polarization vector map of region 1 in NGC 2100 ISP space is over-plotted on a DSS-2 
blue image. \label{n2100region1isp}}
\end{center}
\end{figure}

\begin{figure}
\begin{center}
\includegraphics[scale=0.5]{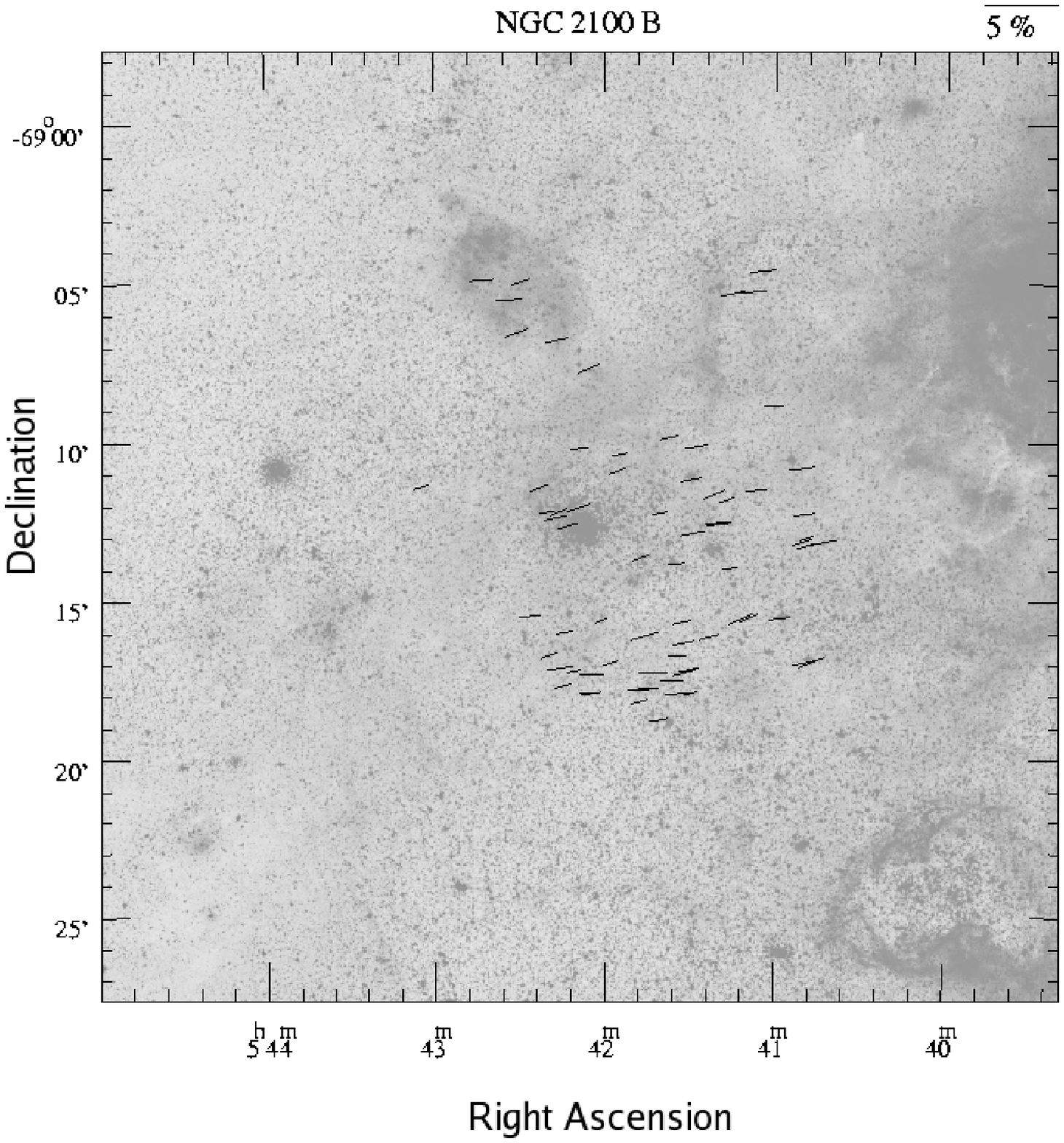}
\caption{The B-band polarization vector map of region 2 in NGC 2100 ISP space is over-plotted 
on a DSS-2 blue image. \label{n2100region2isp}}
\end{center}
\end{figure}

\begin{figure}
\begin{center}
\includegraphics[scale=0.5]{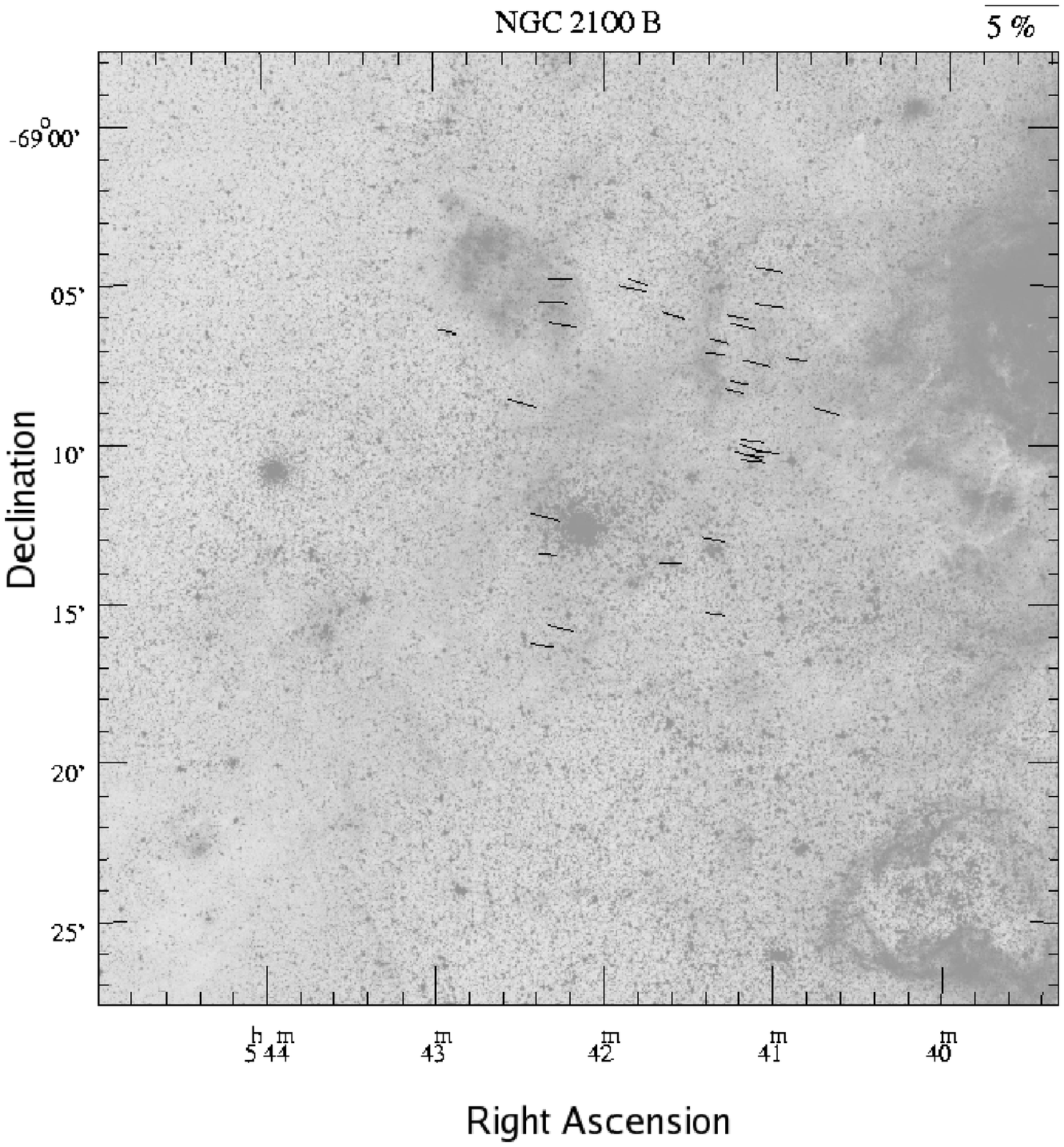}
\caption{The B-band polarization vector map of region 3 in NGC 2100 ISP space is over-plotted 
on a DSS-2 blue image. \label{n2100region3isp}}
\end{center}
\end{figure}

\begin{figure}
\begin{center}
\includegraphics[scale=0.5]{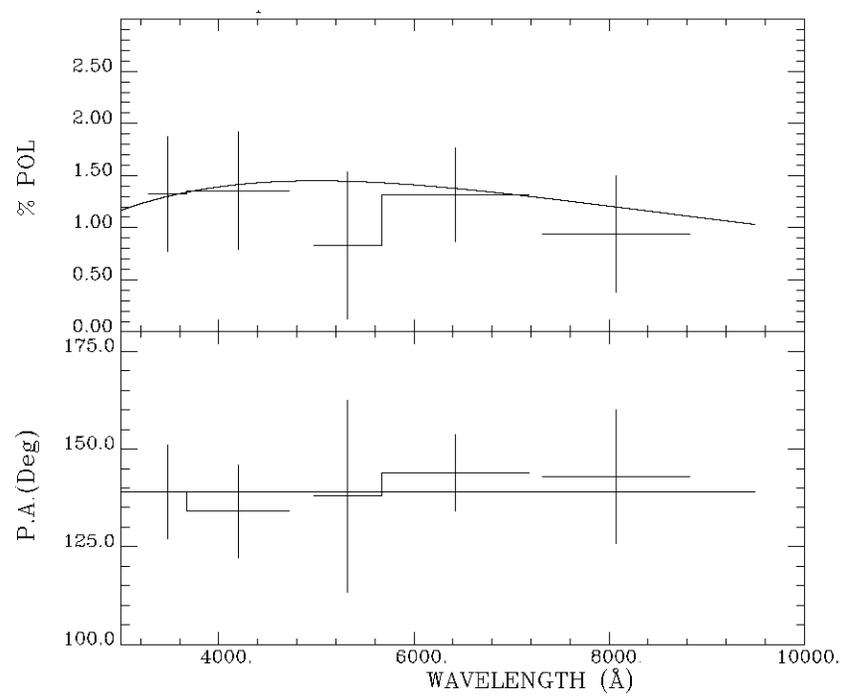}
\caption{The observationally derived initial estimates of the total interstellar polarization towards region 1 of 
NGC 2100 is shown, along with the modified Serkowski-law we deemed best fit the data.  The final ISP 
values for region 1 used in the remainder of this paper were extracted from this Serkowski 
curve. \label{n2100region1serk}}
\end{center}
\end{figure}

We identified three spatial regions in our field of view, corresponding to the region around 
the cluster core, the region to the south of the cluster core, and the region to the north of 
the core,  which displayed unique ISP characteristics.  In Figures 
\ref{n2100region1isp}, \ref{n2100region2isp}, and \ref{n2100region3isp}, we show the B-band 
polarization vector maps of what we have defined as regions 1, 2, and 3 respectively in the 
total ISP along the line of sight to NGC 2100, overlaid on DSS-2 blue images.  Having identified 
these distinct regions, we then extracted total ISP estimates 
following the technique previously described.  We calculated a $\sigma^{-2}$ weighted average 
of all objects in each filter served as an initial ISP estimate, determined the modified Serkowski-law 
parameters which best represented the data, as seen in Figure \ref{n2100region1serk} for region 1, 
and extracted final ISP values (Table \ref{isptable2}) for each region from these Serkowski 
curves.  The locations of the candidate Be stars in NGC 2100 were then correlated to these 3 ISP
regions to determine the ISP correction each should receive.  We found NGC 2100:KWBBe 102, 
630, and 712 resided in area 2, NGC 2100:KWBBe 535, 797, and 1033 resided in area 3, 
and the rest of the candidate Be stars resided in area 1.

\section{Intrinsic Polarization} \label{intrinpolsection}

We subtracted the ISP from the observed polarization to isolate any intrinsic polarization 
components present.  The location of candidate Be stars in our fields of view were identified by 
careful correlation with literature coordinates and finder charts.  Using several PCCDPACK routines, 
we meticulously examined the extracted polarization for each of the candidate Be stars in our sample 
to search for undesired contamination by a) the ordinary or extraordinary images of nearby 
objects and b) cosmic ray hits or uncorrected bad pixels.  The total polarization of all 
candidate Be stars which did not suffer from these contamination issues are tabulated in Table 
\ref{totpol}, while the intrinsic polarization of these objects are tabulated in Table \ref{inpol}.

\begin{figure}
\begin{center}
\includegraphics[scale=0.9]{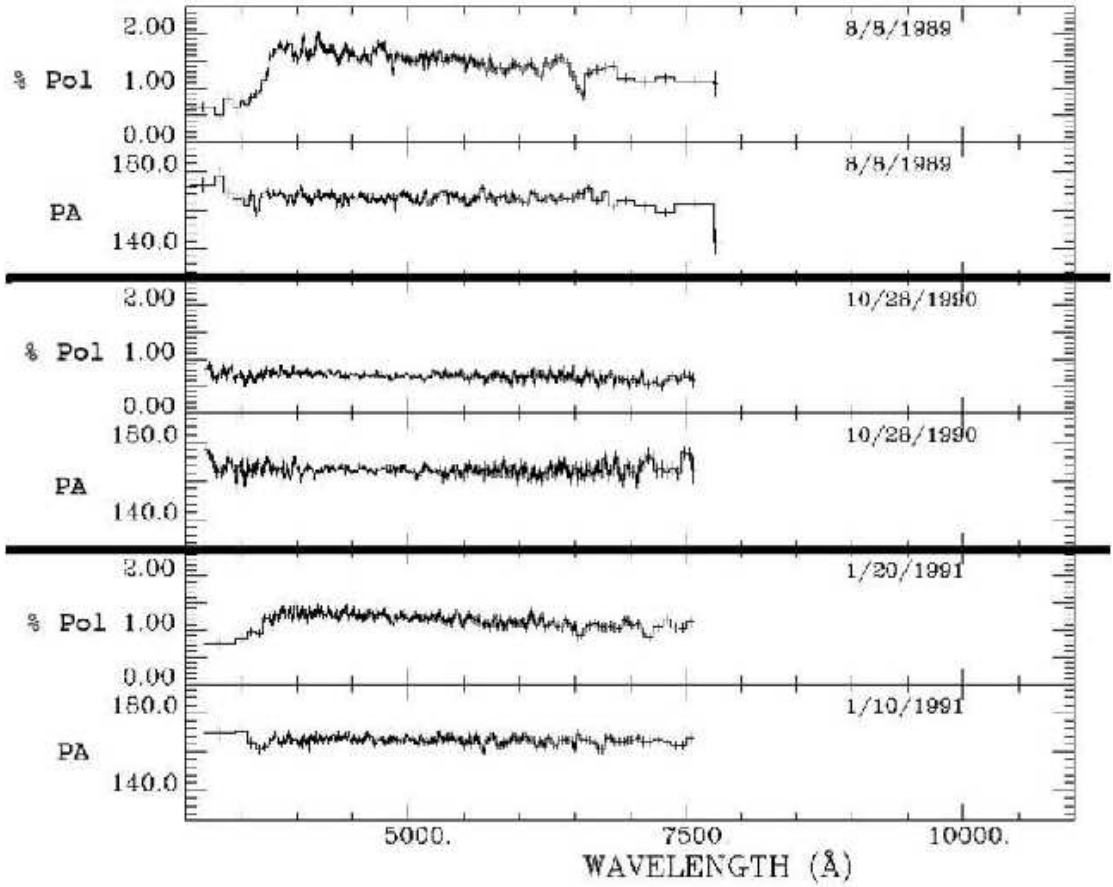}
\caption{The time variability of the intrinsic polarization of the known Galactic classical 
Be star pi Aquarii is shown in these multi-epoch data from the HPOL spectropolarimeter.  
During the 1989 and 1991 observations, the inner disk density 
was sufficiently high such that the effects of pre- and post-scattering absorption by hydrogen 
in the disk is seen in the appearance of a polarization Balmer jump and a ``saw-tooth'' like 
wavelength dependence.  The inner disk density was sufficiently low in the 1990 observation 
such that the wavelength independent electron scattering signature was not noticeably 
altered by a hydrogen opacity signature.  We have examined our intrinsic imaging polarization
observations of candidate Be stars to search for the polarimetric signatures found in this 
figure. \label{pbopiaqr}}
\end{center}
\end{figure}

As discussed in the Introduction, electron scattering in the pure gas disks 
of classical Be stars will polarize a small fraction of stellar photons, producing a well known 
wavelength dependent intrinsic polarization signature.  We emphasize that as polarization is a 
vector quantity, identifying these signatures a) is only possible via inspection of intrinsic 
polarization data; and b) requires either the \textit{simultaneous} inspection of both the 
wavelength dependence of the polarization magnitude \textit{and} position angle, or alternatively 
inspection of the behavior of the intrinsic data on a Stokes Q-U diagram.  An observed 
wavelength independent intrinsic polarization magnitude and position angle across the 
entire UBVRI wavelength regime is the expected signature of pure electron scattering, such as that 
expected from a gaseous classical Be disk.  Moreover, pre- or post-scattering
absorption will superimpose the wavelength dependent signature of 
hydrogen opacity on this signal, creating a ``saw-tooth'' polarization signature \citep{bj00b}, if 
enough absorption events occur.  Specifically, this ``saw-tooth'' signature includes the presence of 
abrupt jumps in the magnitude of intrinsic polarization at the Balmer and Paschen limits, as well as a 
wavelength independent intrinsic polarization position angle across the entire UBVRI wavelength regime.  
Examples of these two signatures of classical Be 
circumstellar disks are given in Figure \ref{pbopiaqr}, which presents multi-epoch observations 
of the known classical Be star pi Aquarii as observed by the University of Wisconsin's HPOL 
spectropolarimeter.  

We have contemporaneously analyzed the 
Stokes Q-U diagrams of all candidate Be stars' total and intrinsic polarization components to 
search for evidence of these signatures.  We have developed a conservative 4-point 
classification scale to rank the likelihood that candidate Be stars are truly classical Be stars.  Results
of this classification are summarized in Table \ref{starclass} for individual objects 
and in Table \ref{polsummary} for the net results of entire clusters. The design of our classification 
system emphasized the identification of objects exhibiting a ``saw-tooth'' intrinsic polarization 
signature and a pure electron scattering signature across the \textit{entire} UBVRI wavelength regime.  Specifically, we defined our classification system by:

\begin{itemize}
\item \textbf{Type-1:} Objects which are \textit{most likely} classical Be stars;
\item \textbf{Type-2:} Objects whose polarimetric properties are \textit{not inconsistent} with those 
expected from classical Be stars;
\item \textbf{Type-3:} Objects which are \textit{unlikely} to be classical Be stars; and 
\item \textbf{Type-4:} Objects which are \textit{highly unlikely} to be classical Be stars.
\end{itemize}

\begin{figure}
\begin{center}
\includegraphics[scale=0.8]{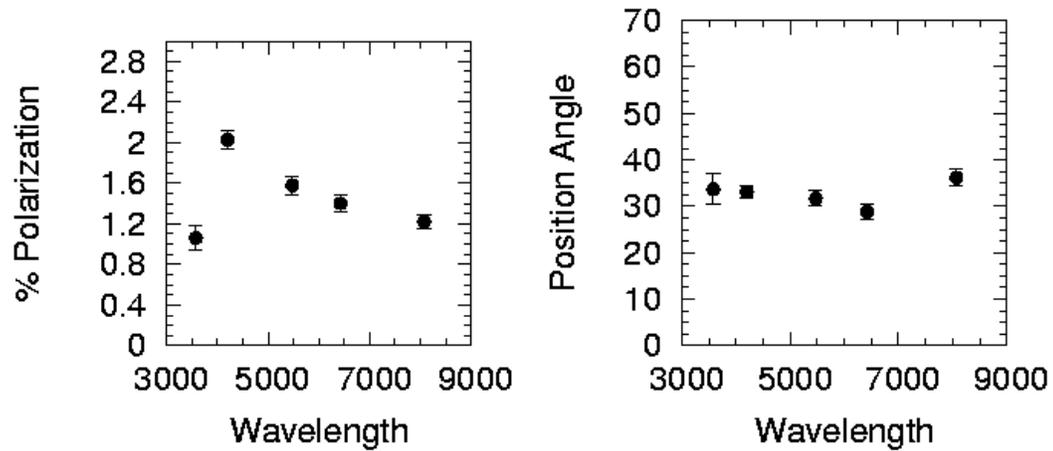}
\caption{The intrinsic polarization of NGC 371:WBBe18 exhibits the ``sawtooth-like'' polarization signature
characteristic of classical Be stars.  Such objects were assigned a designation of type-1 to indicate
they are definitely classical Be stars. \label{polbal}}
\end{center}
\end{figure}

Stars in our sample which displayed, to 
within 3 $\sigma$, a ``sawtooth-like'' polarization Balmer jump (BJ) along with a wavelength independent
polarization position angle across the entire range of available data (see Figure \ref{polbal}) were 
assigned a designation of type-1.  We claim that all type-1 objects are most likely classical Be stars.  
Note that we offer a detailed discussion of other objects which could exhibit somewhat 
similar behavior over \textit{part} of the UBVRI wavelength regime in 
Section \ref{bevshaebe}.  However, as we discuss in Section \ref{nircolors}, analysis of ancillary 
data demonstrates that it is highly dubious that stars classified as Type-1 polarimetric sources are 
anything other than classical Be stars. 

\begin{figure}
\begin{center}
\includegraphics[scale=0.8]{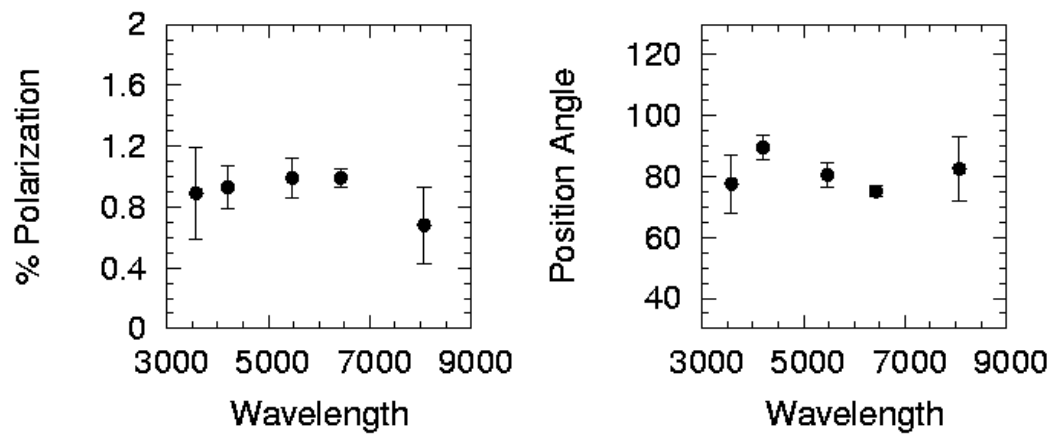}
\caption{The intrinsic polarization of NGC 371:WBBe21 clearly exhibits a wavelength independent 
electron scattering polarization 
signature.  Such objects were assigned a designation of type-2 to indicate they are not
inconsistent with being classical Be stars. \label{poles}}
\end{center}
\end{figure}

We assigned a designation of type-2 to objects whose polarimetric properties were not 
inconsistent with that expected from a classical Be star-disk system.  As previously discussed, 
classical Be disks of sufficiently low density will not leave an imprint of hydrogen opacity in their
intrinsic polarization signals; hence, stars which exhibited, to within 3$\sigma$, a wavelength 
independent electron scattering (ES) polarization signature (see Figure \ref{poles}) received 
a designation of type-2.  A small number of objects exhibited a nearly wavelength independent 
polarization magnitude along with a minor wavelength dependence in their polarization 
position angles; we designated these objects as type-2 as we believe such stars most likely are exhibiting
ES signatures modified by a slight under- or over-correction of their ISP components.  We can not rule out 
that this minor wavelength dependence might also be produced by the additional presence of an optically 
thin dust disk, as present in post-main-sequence B[e] stars \citep{mag92,mel01}.  However, 
as we show in Section \ref{nircolors}, analysis of ancillary data demonstrate that most type-2 
objects which we claim exhibit electron-scattering (ES) signatures are likely to be classical Be stars, 
and substantially less likely to be systems characterized by composite gas plus dust disks 
(i.e. Herbig Ae/Be, B[e] 
stars).  Finally, we 
assigned all objects which appeared to be intrinsically unpolarized ($<$ 0.3\% polarization), 
to within 3$\sigma$, a type-2 designation as this is the expected signature of pole-on or nearly 
pole-on classical Be stars.  We caution the reader that we are not 100\% certain that these 
unpolarized objects are classical Be stars as: a) stars without gaseous disks, spuriously detected as 
excess H$\alpha$ emitters on 2-CDs, will also exhibit zero net intrinsic polarization and b) the 
noise present in the observations of fainter targets may preclude us from clearly identifying 
them as ``contaminants''. 

\begin{figure}
\begin{center}
\includegraphics[scale=0.8]{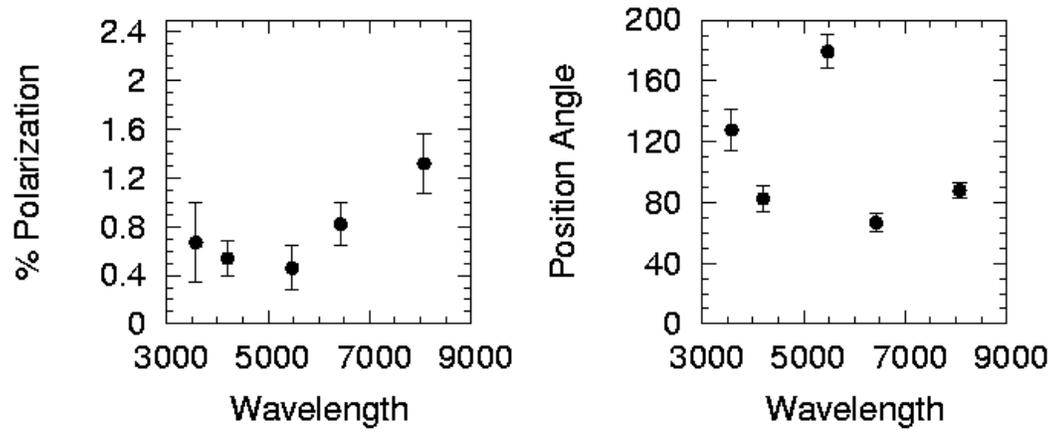}
\caption{The intrinsic polarization of NGC 371:WBBe47 is not consistent, to 
within 3 $\sigma$, with that expected from a classical Be star.  Such objects would receive a 
designation of type-3 or type-4, depending on the severity of their deviation from any of the expected
polarimetric signatures.  These objects (especially those which receive a type-4 designation)
 provide strong evidence that
the 2-CD technique can mis-identify classical Be star-disk systems. \label{polunlikely}}
\end{center}
\end{figure}

We assigned the designation of type-3 to type-4 to objects whose polarization, to within 3$\sigma$,
appeared to be inconsistent with the aforementioned signatures expected from a 
classical Be star-disk system (see e.g. Figure \ref{polunlikely}).  We briefly discuss some of the 
major types of atypical polarimetric signatures we observed.  

In Table \ref{starclass}, we assigned a designation of type 3-4 to several objects 
whose intrinsic polarization exhibited signs of a 90$^{\circ}$ position angle
reversal (Figure \ref{n2100intrin111}), which is a signature 
of a dusty bipolar nebula geometry (see Section \ref{bevshaebe}; see also \citealt{sch92,wup92}).  
From an inspection of the Stokes Q-U diagram of such an 
object (Figure \ref{n2100intrin111}), it is clear that even if our initial ISP correction was grossly
miscalculated, the wavelength dependent polarization would still be inconsistent with that expected
from classical Be stars.  While it would be interesting 
to further probe the circumstellar environments of such objects with follow-up investigations, 
for the purposes of this paper we merely remark that they are unlikely to be classical Be stars.

\begin{figure}
\begin{center}
\includegraphics[scale=0.8]{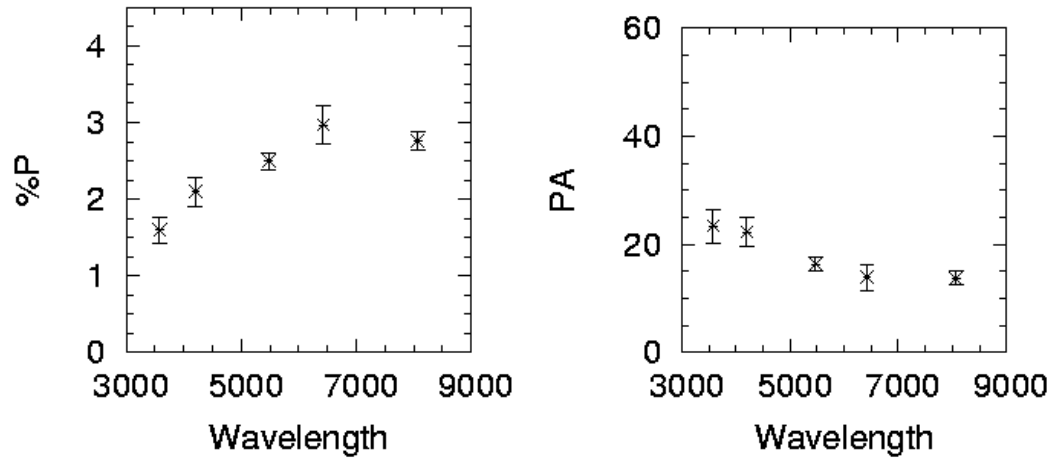}
\caption{The intrinsic polarization 
of NGC 1948:KWBBe 246 follows a complex wavelength dependence.  The optical polarization appears to decrease and position angle increase at short wavelengths.  Objects which have dusty plus gaseous disks, 
such as HD 45677 \citep{wup92}, exhibit similar optical polarimetric behavior, suggesting that 
NGC 1948:KWBBe 246 and similar objects in our sample might also be characterized by such disks.  
Alternatively, the wavelength dependence of these data also resemble a Serkowski-like behavior \citep{ser75}, which as discussed in Section 4 might indicate that some these types of objects 
are characterized by abnormal interstellar dust conditions.   \label{n1948serklong}}
\end{center}
\end{figure}

\begin{figure}
\begin{center}
\includegraphics[scale=0.8]{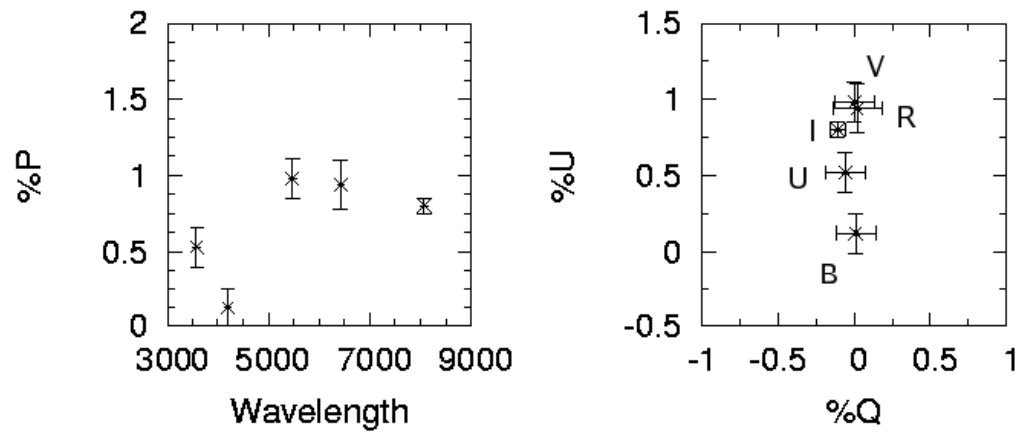} 
\caption{The wavelength dependent intrinsic
polarization of NGC 2100:KWBBe 111 clearly is not consistent with that expected from a classical Be star.  
We suggest that this object might be a dust-disk system. \label{n2100intrin111}}
\end{center}
\end{figure}

Several of our candidate Be stars exhibited very large (1-3.5\%), complex intrinsic polarization 
signatures (see Figure \ref{n1948serklong}).  Given the steep drop in their polarization magnitude at 
short optical wavelengths, along with subtle indications of a corresponding position angle rotation, it 
is possible that these objects are characterized by dusty bipolar nebulae whose polarization signature 
exhibits a position angle reversal at UV wavelengths, similar to HD 45677 \citep{wup92}.  Alternatively, 
the polarization of these objects could be interpreted as following a Serkowski-like wavelength 
dependence characterized by a large $\lambda_{max}$ value.  As our data have already been corrected 
for the average ISP associated with each of our clusters, this latter interpretation would require these 
objects to be situated in region of patchy dust, likely populated by larger grains given the 
long wavelength values of $\lambda_{max}$, i.e. $>$ 7000 \AA\ in 
Figure \ref{n1948serklong} \citep{rod97, whi92}.  \citet{ser75} described the expected 
magnitude of interstellar polarization by the formulation $3 E_{B-V} \le P_{max} \le 9 E_{B-V}$; hence, 
an additional reddening of 0.3 - 1.0 E${(B-V}$ would be required to produce the measured 
p$_{max}$ of these objects, $\sim$3\%.  The observed colors of at least some of these objects (NGC 1948:KWBBe 98, V-I = 0.67; Keller et al. 1999) is likely sufficient to 
produce the amount of purported additional ISP, although other objects (NGC 1948:KWBBe 246, V-I = 0.05; Keller et al. 1999) clearly do not meet this criteria.  While follow-up optical or infrared spectroscopy of 
these anomalous candidates would help to determine whether they exhibit a dusty bipolar nebula 
geometry or are located in a region of patchy dust, for the purpose of this paper we merely stress that it is 
clear that these objects are not likely to be classical Be stars.

We now offer discussion of the polarimetric properties of candidate Be stars
in individual clusters.  Recall that all candidates which exhibited evidence of contamination, 
predominantly from nearby neighbors, have been excluded from our analysis.  We further note 
that the effective limiting magnitude of our polarimetric dataset was lower (i.e. brighter) than that 
considered in WB06.  

\subsection{LMC Clusters}

\subsubsection{NGC 1818}

\citet{kel99} identified 40 candidate Be stars associated with the LMC cluster
NGC 1818 and its surrounding field.  The cluster is densely populated, and because of image
overlap issues, we were only able to retrieve polarimetric information for 18 of these
candidate Be stars.  4 of these 18 (22\%) showed intrinsic polarization
Balmer jump (BJ) signatures, NGC 1818:KWBBe 69, 82, 137, 243.  We classified 12 of the 18 
candidates (67\%) as type-2 objects, and found 5 of the type-2 objects 
(i.e. 5 of the 18 stars, 28\%, in the total population) exhibited clear evidence 
of an electron scattering (ES) polarization
signature.  2 of the 18 candidates (11\%), NGC 1818:KWBBe 47 and 381, were deemed 
unlikely to be classical Be stars.  While the B, V, R, and I filter polarization of NGC 1818:KWBBe 47 
were consistent with an ES signature, the U filter exhibited a significant position angle rotation which 
was inconsistent with an ES origin.  Although noisy, the observation of 
NGC 1818:KWBBe 381 also did not follow an ES wavelength dependence, prompting us to assign it an 
``unlikely Be star'' designation.

\subsubsection{NGC 1948}

\citet{kel99} identified 27 candidate Be stars associated with the cluster
NGC 1948 and its nearby field.  We were able to extract polarimetric 
information on 22 of these candidates from our data set.  6 of these 22
candidates (27\%) exhibited a polarization Balmer jump, 
NGC 1948:KWBBe 71, 75, 102, 153, 172, and 240.
We classified 12 of the 22 candidates (55\%) as type-2 objects, and found 6 of the type-2 
objects (i.e. 6 of the 22, 27\%, in
the total population) exhibited a definite electron 
scattering polarization signature, NGC 1948:KWBBe 62, 92, 101, 157, 326, and 790.  
We note that NGC 1948:KWBBe 92 does show a hint of a small polarization Balmer jump; however, we 
opted to describe its polarization more conservatively, i.e. as having an electron scattering signature.    
We found 4 of the 22 stars (18\%) had polarimetric properties which suggested that they were unlikely to
be classical Be stars. As previously discussed, 3 of these stars, NGC 1948:KWBBe 98, 183, and 246 
exhibited intrinsic polarization signals which seemed to follow a Serkowski-like wavelength dependence. 
NGC 1948:KWBBe 91 was also deemed unlikely to be a classical Be star due to the 
wavelength dependent nature of its polarization position angle.

\subsubsection{NGC 2004}

\citet{kel99} identified 67 candidate Be stars associated with the LMC cluster
NGC 2004 and its surrounding field.  We extracted polarimetric information
for 43 of these candidates, and found that 9 of the 43 (21\%), namely
NGC 2004:KWBBe 50, 87, 96, 103, 106, 152, 211, 347, 377, showed an intrinsic
polarization Balmer jump.  28 of the 43 candidates (65\%) were classified as type-2 objects,  
and note that 10 of these 28 type-2 objects (i.e. 10 of the 43, 23\%, 
in the total population), NGC 2004:KWBBe 91, 203, 276, 323, 344, 441, 624, 717, 1175, and 1421 exhibited an electron scattering polarization signature.  We found 6 of the 
43 (14\%) candidates in NGC 2004 were unlikely to be classical Be 
stars based on their intrinsic polarization.  Of these objects, we note that the 3 filters 
of polarization extracted for NGC 2004:KWBBe 1315 show suggestive evidence of a 
polarization position angle flip, a feature which is not expected in classical Be star-disk
systems.

\subsubsection{LH 72}

WB06 identified 50 candidate Be stars in the LMC cluster LH 72, and
designated 11 of these detections as tentative.  We were able to obtain
polarimetric information for 34 of these stars: 1 of the 34 (3\%), 
LH 72:WBBe 5, showed a polarization Balmer jump and is most likely a 
bona-fide classical Be star.  We classified 22 of the 34 (65\%) stars as
type-2 objects, and found 5 of these 22 type-2 
objects (i.e. 5 of the 34, 15\%, of the total population), LH 72:WBBe 13, 15, 26, 27, and 33 
exhibited an electron scattering 
polarization signature.  We suggest that 11 of the 34 (32\%) of candidates in LH 72 
are unlikely to be classical Be stars based on their intrinsic polarization properties.
Of these unlikely Be stars, we found that the intrinsic polarization of LH 72:WBBe 9 
followed a Serkowski-like wavelength dependence, with a $\lambda_{max}$ value $>$ 
6000 \AA.  We extracted polarimetric information for 7 of the 11 stars 
designated as possible candidate Be stars by WB06.  
5 of these 7 stars were classified as type-2 objects, while 2 of the 7 
appear unlikely to be classical Be stars based on their intrinsic polarization signatures.

\subsubsection{NGC 1858}

WB06 identified 39 candidate Be stars in the LMC cluster NGC 1858, and we 
were able to extract polarimetric information for 27 of these 39 stars.  3 of the
27 (11\%), NGC 1858:WBBe 6, 9, and 20, exhibited polarization Balmer jumps.
We classified 13 of the 27 (48\%) as type-2 objects, and found 2 of these 13 type-2 stars 
(i.e. 2 of the 27, 7\%, of the total population) exhibited an electron scattering polarization 
signature. 11 of the 27 (41\%) candidates in
NGC 1858 appear unlikely to be classical Be stars based on their polarimetric 
signatures.  As previously discussed, 2 of these unlikely Be stars, NGC 1858:WBBe 3 and 12
had an intrinsic polarization which follow a Serkowski-like wavelength dependence, characterized 
by a long $\lambda_{max}$ value.  4 of the 39 photometrically identified candidate Be stars 
in NGC 1858 were judged to be 
possible detections in WB06: we were able to detect 3 of these 4 
polarimetrically.  We classified 2 of the 3 to be type-2 objects, and the remaining object, 
NGC 1858:WBBe 9, as a type-1 object.

\subsubsection{NGC 2100}

\citet{kel99} identified 61 candidate Be stars associated with the LMC cluster
NGC 2100.  We were able to extract polarimetric information for 35 of the
61 candidates (57\%).  8 of the 35 stars (23\%), NGC 2100:KWBBe 79, 97,
436, 619, 635, 705, 770, and 797, exhibited polarization Balmer jumps and hence are most likely to be 
classical Be stars.  20 of the 35 stars (57\%) were classified as type-2 stars, and we found 16 of 
these 20 type-2 stars (i.e. 16 of the 35, 46\%, of the total population) exhibited an
electron scattering intrinsic polarization signature.  We suggest that 7 of the 35 (20\%) 
candidates in NGC 2100 are unlikely to be classical Be stars based on their observed 
intrinsic polarization signatures.  Note that the intrinsic polarization of 
NGC 2100:KWBBe 321 seems to follow a Serkowski-like wavelength dependence, while the
unique intrinsic polarization signatures of NGC 2100:KWBBe 111 and 219 
(Figure \ref{n2100intrin111}) suggests these 
objects might be dust-disk systems.

\subsection{SMC Clusters}

\subsubsection{NGC 346}

\citet{kel99} identified 48 candidate Be stars in the vicinity of the SMC
cluster NGC 346: we have obtained polarimetric information for 33 of
these objects.  8 of the 33 objects (24\%), NGC 346:WBBe 85, 93, 191, 
236, 374, 445, 468, and 529, exhibited polarization Balmer jumps and hence are most likely
classical Be stars.  We classified 22 of the 33 stars (67\%) as type-2 stars, and
found 7 of these 22 type-2 objects (i.e. 7 of the 33, 21\%, of the total population) 
exhibited a clear electron scattering intrinsic polarization signature.
We suggest that 3 of the 33 candidates in NGC 346 (9\%) are unlikely to be classical Be stars 
based on their polarimetric signatures.  Although not detected via our polarimetric 
survey, we note that NGC 346:KWBBe 13 and NGC 346:KWBBe 200 should not be 
considered to be classical Be stars; the former (NGC 346:KWBBe 13) is the well-known 
Wolf-Rayet/LBV HD 5980 and the later (NGC 346:KWBBe 200) has recently been shown to be 
the fifth known B[e] in the SMC \citep{wis07}. 

\subsubsection{NGC 371}

WB06 identified 129 candidate Be stars in NGC 371.  We have obtained polarimetric
information for 73 of these targets, and found 10 of these 73 stars (14\%), 
NGC 371:WBBe 2, 3, 4, 5, 6, 10, 13, 18, 24, and 31 exhibited polarization
Balmer jumps, indicating they are most likely classical Be stars.  We classified 49 of
the 73 stars (67\%) as type-2 objects, and remark that 6 of these 49 type-2 stars 
(i.e. 6 of the 73, 8\%, of the total population) exhibited a clear electron scattering polarization
signature.  Note that many of the fainter candidate Be stars included in our
74 detections were not observed at high signal-to-noise levels; thus for most
of these objects we can only say that to within 3$\sigma$, their polarimetric
properties are not inconsistent with those of classical Be stars.  We suggest that 
14 of the 73 (19\%) candidates in NGC 371 are unlikely to be classical Be stars 
based on their polarimetric properties.  WB06 suggested that 11 of 
the 130 photometrically identified candidates 
should be viewed as ``possible detections''; however, we were only able to extract 
polarimetric information for 2 of these 11 objects, NGC 371:WBBe 64 and 87.  We 
classified both as type-2 stars.

\subsubsection{Bruck 60}

We were able to extract polarimetric information for 18 of the 60 candidate Be stars 
identified photometrically by WB06.  We found 1 of the 18
(6\%), Bruck 60:WBBe 6, showed a polarization Balmer jump, while we classified
 11 of the 18 stars (61\%) as type-2 objects.  8 of these 11 type-2 
stars (i.e. 8 of the 18, 44\%, of the total population) exhibited an electron scattering 
intrinsic polarization signature.  We suggest that
6 of the 18 (33\%) candidates in Bruck 60 are unlikely to be classical Be stars based on their
polarimetric signatures.  WB06 noted that 5 of the 26 photometrically 
identified candidate Be stars in Bruck 60 
should be considered ``possible detections''.  We were only able to detect 1 of 
these 5 stars polarimetrically, Bruck 60:WBBe 21, and classified it as a type-2 object which 
exhibited clear evidence of an electron scattering polarization signature.

\subsubsection{NGC 456}

23 candidate Be stars in NGC 456 were identified in the photometric survey of WB06, and
we were able to extract polarimetric information for 14 of these 23 candidates.  Although 
we did not observe any objects with polarimetric Balmer jumps, we did classify 9 of the 14 stars
(64\%) as type-2 objects.  1 of these 9 type-2 stars (i.e. 1 of the 14, 7\%, of the total population), 
NGC 456:WBBe 1, exhibited an electron scattering polarization signature.  
We suggest that 5 of the 14 (36\%) candidates in NGC 456 are unlikely to be classical Be stars 
based on their observed intrinsic polarization components.  WB06 suggested 1 
of NGC 456's 23 photometrically identified candidate Be 
stars should be viewed as a ``possible detection'': we marginally detected this object, 
NGC 456:WBBe 21, and suggest that it is unlikely to be a classical Be star. 

\subsubsection{NGC 458}

30 candidate Be stars were photometrically identified in NGC 458 by WB06, and
we were able to extract polarimetric information for 10 of these 30 candidates.  None of
the 10 detected objects exhibited polarimetric Balmer jumps or clear evidence of having an
electron scattering signature.  We classified 6 of the 10 stars (60\%), and suggest that 
4 of the 10 (40\%) are unlikely to be classical Be stars.  We were able to extract 
polarimetric information for 1 of the 2 initially identified as ``possible 
detections'' by WB06, NGC 458:WBBe19.  We classified this star as a type-2 object; 
furthermore, we found it exhibited suggestive evidence of a small polarimetric Balmer jump, 
although the data were very noisy.

\subsubsection{NGC 330}

\citet{kel99} identified 76 candidate Be stars in the vicinity of NGC 330 and we
were able to extract polarimetric information for 41 of these objects.  While we did not
observe any of the detected candidates to have a polarimetric Balmer jump, we did classify  
24 of the 41 (59\%) candidates as type-2 stars, and note that 9 of these 24 type-2 
objects (i.e. 9 of the 41, 22\%, of the total population) exhibited electron scattering 
polarimetric signatures.  We suggest that 17 of the 41 (41\%) candidates are unlikely to 
be classical Be stars based on their intrinsic polarimetric signatures.

\section{Discussion}

\subsection{Analysis of Intrinsic Polarization Statistics}

It is clear from Table \ref{polsummary} that the percentage of SMC and LMC candidate Be stars which 
exhibit polarization Balmer jumps, electron scattering signatures, or were deemed unlikely
to be classical Be stars varies significantly from one cluster to the next.  To better assess 
the global trends of our dataset, we extracted simple statistical mean and median values from
Table \ref{polsummary}, as summarized in Table \ref{polsummary2}.    From Table \ref{polsummary2}, we 
see that at 
least 25\% of our dataset is populated by objects which appear unlikely to be classical Be stars.  
The presence of such a large number of likely ``contaminants'' is interesting, as it confirms one of the
initial hypotheses of this project: it is dangerous to assume, a-priori, that all objects identified as 
excess H$\alpha$ emitters via 2-CD surveys are classical Be stars.  We thus suggest caution should 
be exercised when attempting to ascertain the role evolutionary age and/or metallicity play in the development 
of the Be phenomenon purely via the analysis of 2-CD data.  

Assuming these objects are truly ``contaminants'', we recalculated the statistics of our intrinsic 
polarization dataset after removing these objects from consideration.  The results of this 
exercise are listed in Table \ref{polsum3}.  The median prevalence of polarization Balmer jumps in 
our SMC/LMC dataset is typically between 20-25\% (Table \ref{polsum3}).  Polarization Balmer jumps 
appear to be significantly less prevalent in 
our SMC clusters (4\%) than in our LMC clusters (25\%); however, this statistic is strongly influenced by the 
null detection of such signatures in three SMC clusters.  

Since 1989, the HPOL spectropolarimeter \citep{wol96} mounted on the University of Wisconsin's Pine Bluff
Observatory (PBO) has been monitoring the optical spectropolarimetric properties of a sample of 73 known 
Galactic Be stars, and it provides a statistically significant
polarimetric dataset which we will use to aid the interpretation of our results.  As yet, most of the $\sim$800  
HPOL observations of classical Be stars have not had their ISP components identified and 
removed; hence, we can not
readily compare their intrinsic polarization characteristics to our data.  However, we can determine the 
prevalence of polarization Balmer jumps in the HPOL database.  Data obtained prior to 1995 have 
been compiled in catalog form by \citet{bjo00}, and observations characterized by a polarization Balmer jump
have already been flagged in the catalog.  We have similarly analyzed all observations obtained from 1995-2004 to 
identify the presence of polarimetric Balmer jumps in these data.  We then compiled all HPOL observations 
of individual classical Be stars into two categories: a) classical Be stars which had \textbf{never} 
exhibited a polarimetric Balmer jump in any PBO observation; and b) classical Be stars which had 
exhibited a polarimetric Balmer jump in at least one observation.  The results of this classification 
procedure are given in Table \ref{polsum3}.

We found 31 of the 73 (42\%) observed Galactic Be stars exhibited a polarimetric Balmer jump for at least 
part of the time-frame of the HPOL survey.  Amongst early-type Be stars (O9-B5) in the HPOL database, 22 
of 45 (49\%) exhibited polarimetric Balmer jumps.  Note that further restricting the sample to include only 
O9-B3 type Be stars does not appreciably affect the observed prevalence of polarization Balmer jumps, as 17 of 
40 (43\%) exhibited such a signature.  In contrast, the prevalence 
of polarization Balmer jumps does seem to be noticeably lower for later-type (B6-A0) Be stars, as we found
only 8 of 26 (31\%) exhibited evidence of a polarimetric Balmer jump. 

The frequency of polarization Balmer jumps appears to decrease with metallicity: $\sim$45\% of Galactic Be stars in
the HPOL database, $\sim$25\% of LMC Be stars from the present study, and $\sim$4\% of SMC Be stars from the 
present study exhibit polarization Balmer jumps.  We suggest several possible explanations for these results.
\begin{enumerate}
\item 
It is possible that the average disk properties of classical Be stars are fundamentally
different in the low metallicity environments of the SMC and LMC as compared to our Galaxy.  The presence of
intrinsic polarization in classical Be stars, as well as its wavelength dependence, is a function of disk inclination 
angle, disk density, and the effective temperature of disk material.  Thus the observed lower frequency 
of polarization Balmer jumps in our data suggest either that it might be harder to form massive disk systems 
in low metallicity environments or that the average disk temperature is higher in these lower metallicity environments, 
hence decreasing the amount of pre- and post-scattering absorption by neutral H I.
  
\item Our polarimetric observations provided a one time sampling of LMC and SMC candidate Be 
stars, while the HPOL survey obtained a larger number of observations, albeit at quasi-random intervals, of
a population of known Be stars.  Owing to the variable nature of the Be phenomenon, we do not a priori know 
how often any individual Be star will exhibit any of the possible polarization signatures; hence, the different 
sampling errors associated with these two databases might explain at least some of the 
observed differences in the polarization Balmer jump frequencies.

\item The statistics presented in Table \ref{polsum3} were based on removing a significant 
number of objects deemed unlikely to be classical Be stars from consideration, i.e. type-3 and type-4 objects.  
Recall however that our definition of type-2 objects only required their polarimetric properties to be 
not inconsistent with those expected from classical Be stars.   It is possible that a population of 
``contaminants'', either a) unpolarized objects which aren't pole-on or near pole-on classical Be stars; 
or b) non-classical Be stars which were not identified as such due to low data quality, still reside within 
our collection of type-2 objects, hence underestimate our polarization Balmer jump fraction frequency.  
If such a 
scenario is true, it would further suggest that extreme caution should be exercised when assessing the 
nature of stellar populations purely via the use of photometric 2-CDs.

\end{enumerate}

\subsection{Identification of ``Contaminants'' in 2-CD Detected Candidate Be Stars}

WB06 identified a large number of B-type objects having excess H$\alpha$ emission via 
photometric 2-CDs and suggested that these stars be classified as candidate Be stars.  D. Gies (2006, 
personal communication) noted that several candidate Be stars in the WB06 dataset 
had colors redder than that expected from classical Be stars, and suggested that the 
frequency of observed blue (or red) excess H$\alpha$ emitters in the 2-color diagrams presented in 
WB06 increased with the total number of stars of a given color range.  Furthermore, 
he postulated that if these red excess emitters were a) a manifestation of 
extremely young stars with their birth disks still intact, b) red-type giants or supergiants, or 
c) false detections owing from the effects of bright background H II emission; measuring the 
prevalence of these excess emitters could serve as 
an estimate of the number of ``contaminants'' which affect the blue population of 
excess H$\alpha$ emitters, i.e. the candidate Be star population.  

We examined the 2-CD photometry of 6 clusters,  
common to both this survey and that of WB06 and spanning a wide range of 
ages (very young, young, old; WB06)  and metallicities (SMC, LMC), and 
determined the frequency of excess red-type  
H$\alpha$ emitters relative to the total number of red-type stars.  We restricted our analysis such 
that the magnitude of (B-V) colors used to identify candidate Be stars in each cluster (i.e. 
$\sim$ -0.3 $<$ (B-V) $<$ 0.2 in NGC 458, WB06) matched the range of (B-V) colors 
used to study the frequency of red-type excess emitters (i.e. 0.2 $<$ (B-V) $<$ 0.7 for NGC 458).  
Furthermore, we adopted the same (R-H$\alpha$) cutoffs delineating excess emitters from normal 
stars for each cluster as used by WB06.  The results of this exercise 
are tabulated in column 2 of Table \ref{photcontam}.  Interestingly, 
with few exceptions, the frequency of potential ``contaminants'' implied by this technique is 
generally consistent with the minimum rate of contamination derived from our polarization results
(see column 3 of Table \ref{photcontam}, column 6 of Table \ref{polsummary}), i.e. the candidates 
we assigned a Type 4 classification.  Clearly this photometric contamination check can not provide 
a diagnostic on the Be classification of individual objects, unlike our polarimetry; however, 
we suggest that this technique might provide a rough estimate of the role of contaminants in 
cases in which detailed follow-up observations have not been made or are unfeasible.

\subsection{The Evolutionary Status of Excess H$\alpha$ Emitters in Very Young (5-8 Myr) Clusters: Early-Type Classical Be stars or Early-Type Herbig Be stars?}

One of the interesting results presented in WB06 was the 
identification of candidate Be star in clusters 
of age 5-8 Myr, in an abundance similar to the general frequency of the Be 
phenomenon in our Galaxy, $\sim$17\%.  However, it was uncertain based on those results 
whether the detected excess H$\alpha$ emitters a) were true classical Be star-disk systems; 
b) were B-type objects which still possessed remnant star-formation disks, i.e. Herbig Be stars; or c) merely 
had diffuse H$\alpha$ nebulosity coincidentally associated with them.  Our intrinsic polarimetric 
dataset affords us the opportunity to further constrain the evolutionary status of candidate Be stars in 
four clusters of age 5-8 Myr, LH 72, NGC 1858, NGC 346, and NGC 371.

\subsubsection{Intrinsic Polarization Signatures of Dust versus Gas Disks} \label{bevshaebe}

As discussed in the Introduction, Section \ref{intrinpolsection}, and 
references therein, it is well accepted that classical Be stars are characterized by geometrically thin gas disks, and the mechanism responsible for 
producing their intrinsic polarization is electron scattering.  Herbig Be stars are 
more complex systems 
in that their circumstellar environments contain both gas and dust.  The \textit{dominant} 
polarigenic mechanism in Herbig Be stars is well established to be scattering by dust, as outlined in the 
review papers of \citet{bas88}, \citet{gri94} and \citet{wat98} and numerous references therein.  
While Herbig Be and classical Be stars exhibit some similar observational traits, such as the presence of an intrinsic, often variable, polarization component \citep{vra79,qui97,vin02,bjo05,bjo06}, the different scattering 
mechanisms which characterize Herbig Be versus classical Be stars manifest themselves in several  
observational manners.  The magnitude of linear polarization of classical Be stars is typically of 
order 1\% or less \citep{bjo00}, whereas the average ($\sim$3\%) and maximum (14.5\%) polarization 
observed in Herbig Ae/Be stars is clearly much larger \citep{tam05}.  More importantly, the different 
scattering mechanisms which dominate in each of these star-disk systems will produce a different 
wavelength dependence in these stars' intrinsic polarization.  

The characteristic wavelength dependent intrinsic polarization signature of classical Be stars has already been documented in Section \ref{intrinpolsection}.  Herbig Be stars are not known
to exhibit any singular ``characteristic'' intrinsic polarization behavior (see e.g. 
\citealt{vra79,bas88,mey02}); the grain chemistry, grain size distribution, and relative size and 
extent of the gas and dust disks/envelopes of individual Herbig Be systems likely contributes to the 
heterogeneous mix of observed intrinsic polarization.  Some typical intrinsic polarization trends 
are observed in Herbig Ae/Be stars however, most notably rotations (and dramatic 90$^{\circ}$ reversals) 
in the intrinsic polarization position angle \citep{sch92,bjo95,bjo98,mey02}.  We 
remark that several literature works 
which describe and interpret the wavelength dependence of intrinsic polarization from 
Herbig Be stars, and at times 
attempt to draws parallels with the intrinsic polarization behavior of classical Be stars, should be 
viewed with caution.  For example, as noted by \citet{bas88}, both \citet{vra75} and \citet{gar78} employ 
dubious assumptions in their efforts to constrain the interstellar polarization along the line of sight 
to their Herbig Be stars, and as such the wavelength dependence of the ``intrinsic'' polarization 
they report is likely inaccurate.  
Similarly, the discussion of the wavelength dependence of intrinsic polarization of the Herbig Be star 
BD+61 154 in \citet{vra79} (see their Figure 2) ignores the wavelength dependence of the intrinsic polarization 
position angle.  The $\sim$40$^{\circ}$ intrinsic polarization position angle rotation from the B to I filter in 
these data is inconsistent with the expected signature of a pure gas disk system; the wavelength 
dependence of intrinsic polarization from this Herbig Be star is not similar to that observed for 
classical Be stars.

Although dust scattering is the dominant polarigenic agent for Herbig Be stars, it is still possible to 
detect evidence of the gaseous inner disk component of some of these systems via polarimetry
 \citep{vin02,mey02,oud05}.  For example, \citet{mey02} presents intrinsic spectropolarimetry   
of the Herbig Be star MWC349A covering the wavelength range $\sim$5000-10500\AA\ (V,R,I filters), 
which exhibits both line depolarization effects and a Paschen 
polarization jump (see their Figure 7), similar to that observed in the classical Be star zeta Tau \citep{woo97}.  \citet{mey02} suggest electron scattering as the mechanism responsible for producing 
these specific features.  Evidence of the dusty component of MWC349A's composite disk is also evident 
in these data.  Specifically, the 30-40$^{\circ}$ rotation in the intrinsic continuum polarization position 
angle, beginning at $\sim$6200 \AA\ and extending to the blue edge of the available data, $\sim$5000 \AA\, indicates the presence of an additional scattering mechanism which becomes more 
pronounced at shorter wavelengths, i.e. dust scattering.  

Based on the available literature, we are aware of no compelling examples of Herbig Be stars whose 
intrinsic polarization magnitude \textit{and} position angle fully follow the known UBVRI behavior 
of classical Be stars.  While neither typical nor likely, we can not rule out the possibility that one could 
in principle observe a Balmer jump signature in the polarization magnitude of \textit{some} Herbig 
Be stars, as these systems do have composite gas plus dust disks.  However, as 
dust scattering is the dominant polarigenic scattering agent in Herbig Be stars, it is dubious that the entire 
UBVRI intrinsic polarization magnitude and position angle of such systems would exhibit no evidence of a 
non-electron scattering component.  Moreover, as astutely pointed out by our referee, it would be 
even more unlikely to observe one of our ``Type 2 electron scattering (ES) signatures'' in any Herbig 
Be star, unless the dust grain chemistry and size distribution would conspire to produce 
grey scattering.  

Hence, the UBVRI spectral range of the intrinsic polarization data presented in this paper enables 
us to discriminate deviations from a pure electron scattering signature over a broad spectral range.  This
allows us to differentiate Herbig Be stars analogous to MWC 349A, whose gaseous disk dominates a portion of  
the optical polarimetric regime, and Herbig Be stars whose dust disks dominate the optical polarimetric 
regime, from classical Be stars.

\subsubsection{Nebular Origin?}

As noted in WB06, many of the candidate Be stars identified in clusters of age 5-8 Myr reside nearby 
or even within regions characterized by diffuse nebular emission.  We consider whether the observed 
characteristic polarimetric signatures of classical Be stars, both electron scattering and Balmer jump 
polarization features, could merely be artifacts of this background emission.  Such polarization 
signatures are only produced when the electron scattering optical 
depth is $\sim$1.0, i.e., that present in the innermost disk region of classical Be stars.  The density 
of any coincident diffuse H$\alpha$ nebulosity present around candidate Be stars is vastly insufficient 
to create such a polarization signature.  Thus while some of the ``Be star'' detections made by 
H$\alpha$ spectroscopic surveys \citep{maz96} have been shown to be spurious detections due 
to the presence of coincident gas \citep{kel98}, such diffuse nebulosity can not be responsible for creating 
the polarimetric signals found in our dataset.

\subsubsection{Near-IR Colors} \label{nircolors}

Further insight into the evolutionary status of candidate Be stars in our very young (5-8 Myr) clusters 
may be elicited via inspection of their near-IR colors.  It has been shown that the fundamentally 
different composition of classical Be disks (gas) and Herbig Ae/Be disks (gas and dust) will generally 
result in these objects occupying distinctly different locations in (J-H) versus (H-K) near-IR 2-color diagrams \citep{lad92,li94}, although minor instances of overlap are observed.  As such, we cross-correlated all photometrically identified candidate Be stars residing in clusters studied in this paper with the 2MASS \citep{skr06} survey.  Targets for which we were able to extract full JHK photometric data are 
listed in Table \ref{ircolor}.

\begin{figure}
\begin{center}
\includegraphics[scale=0.8]{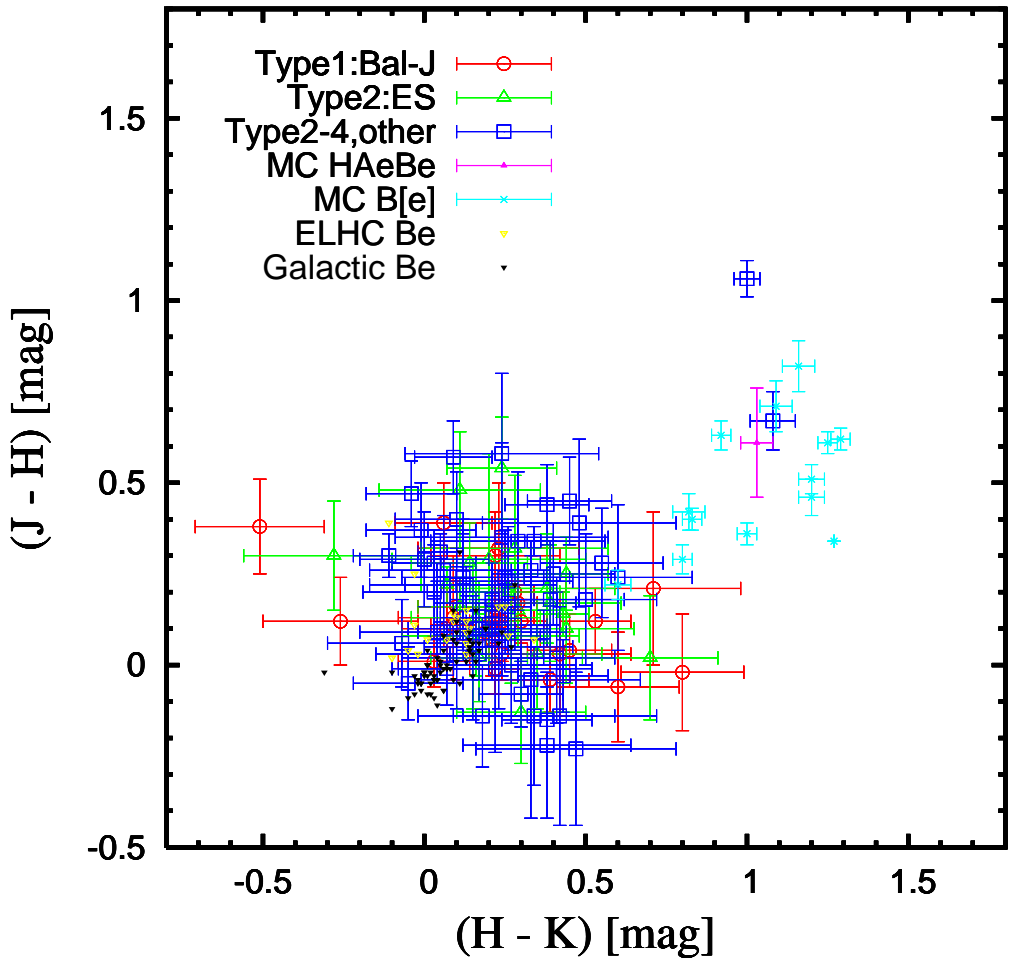} 
\caption{This near-IR 2-color diagram presents all available 2-MASS photometry of candidate 
Be stars residing in clusters investigate in this paper.  ``Type-1:Bal-J'' sources (red circles)  
correspond to our polarimetrically designated Type-1 sources which we assert are most likely
 to be classical Be stars, ``Type-2:ES'' sources (green triangles) correspond to our polarimetrically 
designated Type-2 sources which exhibit electron scattering intrinsic polarimetric signatures and are 
thus likely classical Be stars, ``Type-2-4,other'' sources (dark blue squares) correspond to 
all other candidate Be stars in clusters investigated in this paper regardless of whether 
they were detected via our polarimetric survey or not, ``ELHC Be'' sources (yellow triangles) correspond 
to LMC stars which \citet{dew05} claim are likely to be classical Be stars, ``Galactic Be'' sources (black 
triangles) are 101 known Galactic Be stars tabulated by \citet{dou91}, ``MC B[e]'' (light blue cross) 
correspond to 2-MASS and ground-based \citep{gum95} photometry of known Magellanic 
Cloud B[e] supergiants, and ``MC HAeBe'' (pink triangle) corresponds to the star ELHC-7, which \citet{dew05} assert is a Herbig Ae/Be star which resides in the LMC.  As discussed in 
Section \ref{nircolors}, most of the candidate Be stars which we were able to correlate with 
the 2MASS catalog exhibit colors consistent with those expected from classical Be stars, and 
inconsistent with those expected from composite gas plus dust disk systems (i.e. Herbig 
Ae/Be or B[e] stars). \label{ir2cd}}
\end{center}
\end{figure}

\begin{figure}
\begin{center}
\includegraphics[scale=0.8]{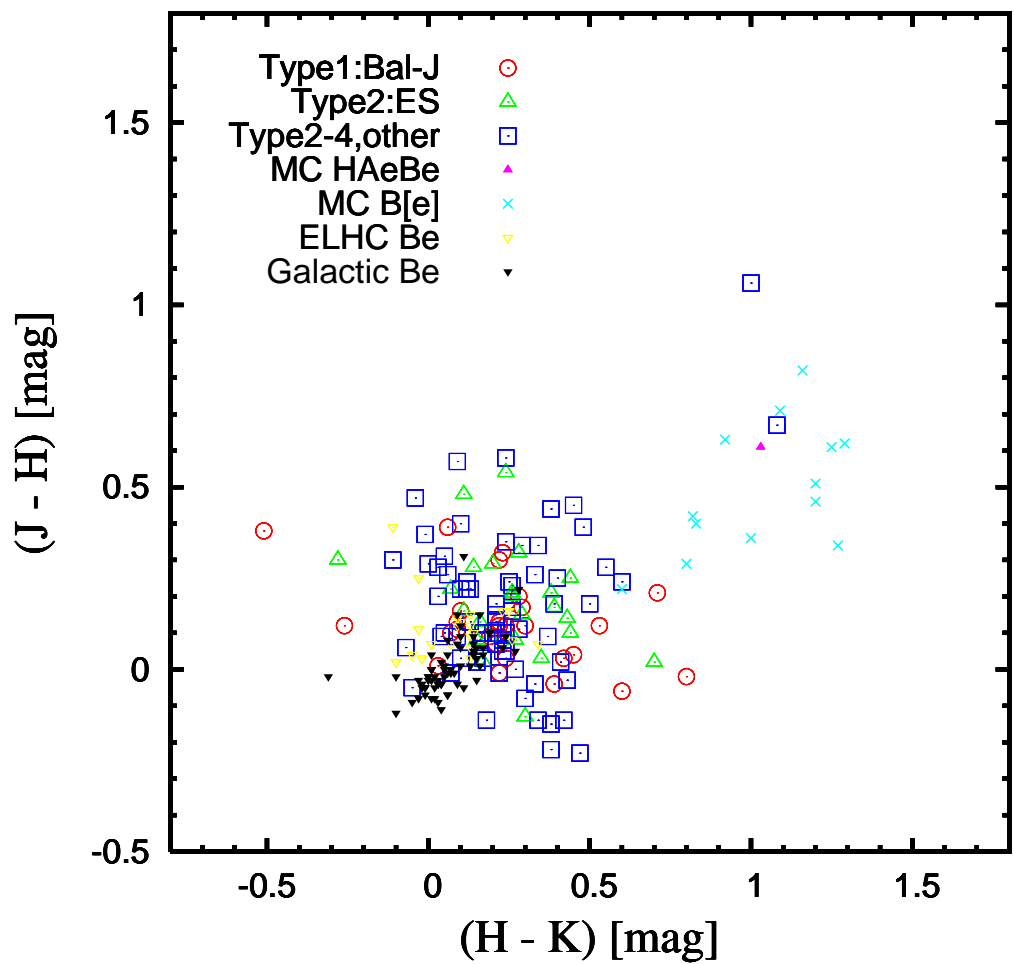} 
\caption{The same data as presented in Figure \ref{ir2cd}, without error bars. \label{ir2cdnoe}}
\end{center}
\end{figure}

We plot the 2MASS JHK photometry of all available sources as a function of 
our polarimetric classifications in Figure \ref{ir2cd} (with error bars) and Figure \ref{ir2cdnoe} 
(without error bars).  Our Type-1 sources 
which exhibit polarimetric Balmer jumps are plotted as red circles, our Type-2 sources which exhibit 
ES polarimetric signatures are plotted as green triangles, and all other candidate Be stars (other Type-2 sources, Type-3 sources, Type-4 sources, and polarimetric non-detections) are plotted as blue squares.  For reference, we have also plotted the near-IR colors of 101 known Galactic Be stars tabulated by \citet{dou91} as filled black triangles, 
21 LMC ``ELHC'' stars which \citet{dew05} suggest are likely to be classical Be stars as yellow 
open triangles, 2MASS and ground-based \citep{gum95} colors of 13 Magellanic Cloud B[e] stars 
known to have dusty disks as light blue crosses, and the star ``ELHC-7'' which \citet{dew05} suggest 
is likely to be a LMC Herbig Ae/Be dusty star-disk system as a pink triangle.

Figures \ref{ir2cd} and \ref{ir2cdnoe} clearly demonstrate that most of the candidate Be stars which we were able to correlate with the 2MASS catalog do lie in a distinctly different region of the JHK  2-color diagram than that occupied by known dustier 
(Herbig Ae/Be or B[e]) systems.  Furthermore, although the Magellanic Cloud candidate Be 
stars' colors exhibit a wide dispersion, likely due in part to the considerable photometric errors of these 
faint sources (Figure \ref{ir2cd}; Table \ref{ircolor}), we find no clear distinction amongst 
the near-IR colors of 
our candidate Be stars as a function of their polarimetric classification.  A small number of candidate Be 
stars marginally overlap with the location of dustier B[e] sources, although these sources 
are within 3-$\sigma$ of the mean locus of classical Be colors.  Clearly better quality near-IR photometry 
of these sources should be pursued to better elucidate their evolutionary status.  Two candidate Be stars 
clearly exhibit near-IR colors consistent with dusty disk systems, NGC 346:KWBBe 200 and NGC 456:WBBe 7.  NGC 346:KWBBe 200 was not detected in our polarimetric survey owing to contamination from a nearby 
source; however, \citet{wis07} has recently published optical spectroscopic, near-IR photometric, and 
IR photometric observations of the target and concluded it is a B[e] supergiant, not a classical 
Be star.  NGC 456:WBBe 7 was observed in our polarimetric survey and assigned a classification of Type-3.5.  Although noisy, our intrinsic polarization data exhibited marginal evidence of a position angle 
rotation, which could be indicative of a dusty envelope.  The star's near-IR colors seems to verify this 
tentative interpretation.

\begin{figure}
\begin{center}
\includegraphics[scale=0.8]{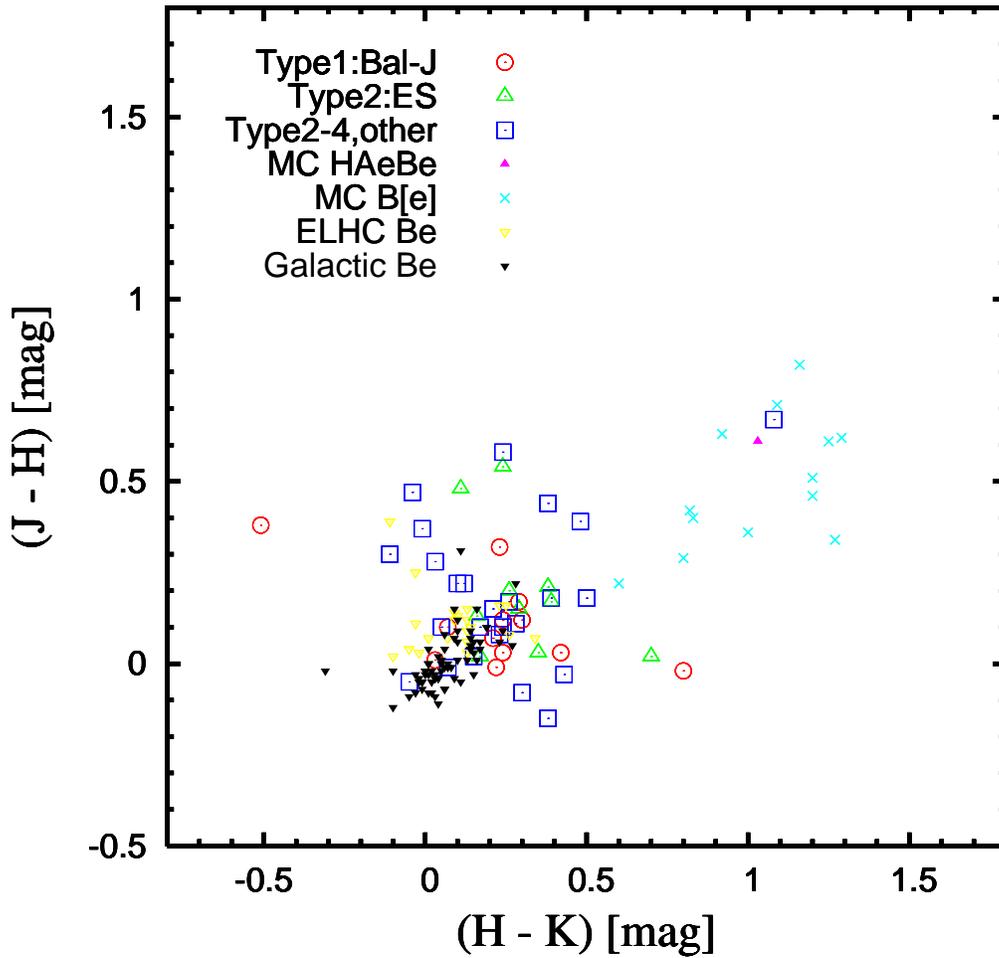} 
\caption{The subset of data presented in Figure \ref{ir2cdnoe}, corresponding to candidate Be 
stars residing in ``very young'' clusters of age 5-8 Myr, is presented.  As discussed in 
Section \ref{nircolors}, the data presented in this Figure and Figure \ref{ir2cdnoe} exhibit 
similar properties.  Regardless of polarimetric classification type, all candidate Be stars in clusters 
of age 5-8 Myr appear consistent with the expected colors of classical Be stars and inconsistent with 
the observed colors of composite gas plus dust systems (i.e. Herbig Ae/Be, B[e] stars).  We 
interpret these data as additional evidence that many of the photometrically identified candidate 
Be stars in clusters 5-8 Myr old by WB06 are classical Be stars and not pre-main-sequence 
Herbig Be stars.   \label{ir2cdvy}}
\end{center}
\end{figure}

For comparison, we also plotted the same near-IR colors for only candidate Be stars residing in our very young (5-8 Myr) clusters in Figure \ref{ir2cdvy}.  Comparison of Figures \ref{ir2cdnoe} and \ref{ir2cdvy} 
reveal no distinctive differences; rather, it is quite clear that the vast majority of candidate Be stars in these 
very young clusters have near-IR colors consistent with those expected from classical Be stars and 
notably inconsistent with those expected from composite gas plus dust disk systems 
(Herbig Ae/Be and B[e] stars).  Similar to 
Figure \ref{ir2cdnoe}, Figure \ref{ir2cdvy} reveals that there is no clear color difference amongst our various 
polarimetric classifications.  We interpret these results as additional 
evidence that the many of the candidate 
Be stars identified in clusters 5-8 Myr old by WB06 are classical Be stars, and not pre-main-sequence 
Herbig Be stars.

\subsubsection{Natal Disk Clearing Time}

The clearing time-scale of natal star formation material (gas and dust) will influence whether it is 
possible or likely, a-priori, to observe early-type Herbig star-disk systems in clusters of age 5-8 Myr. Amongst the less massive Herbig Ae and T 
Tauri stars, the median lifetime of inner optically thick accretion disks is 2-3 Myr, and although 
exceptions do exist, there is little to no H- and K-band evidence of primordial disks beyond 
a median stellar age of 5 Myr \citep{hil05}.  The typical natal disk dissipation for more massive stars 
is less well known.  From a detailed study of the young cluster NGC 6611, \citet{dew97} 
suggested that the clearing of natal disk material around the more massive stars in this cluster 
was typically 0.1 Myr or less.  Similarly, \citet{pog06} recently suggested that B0e star HD 53367 is a 
classical Be star, based in part on an observed episodic loss of its disk material, a phenomenon 
known to characterize many classical Be stars \citep{por03}.  \citet{pog06} also identified a 
4-5 M$_{solar}$ pre-main-sequence binary companion which, given the pre-main-sequence 
evolutionary tracks of \citet{pal93}, suggests the system's age is $<$ 0.8 Myr and indicates the natal 
gas plus dust disk of the B0e star HD 53367 must have dissipated at least on this time-scale.  Although 
clearly not conclusive, these results do suggest that it is reasonable to expect that most of the 
natal disks will have been cleared from the early B-type stars in our clusters of age 5-8 Myr.

\subsubsection{Summary: Most Candidate Be stars in Clusters 5-8 Myr Old Are Classical Be Stars} \label{polconcl}

WB06 and \citet{kel99} identified a large population of B-type stars in clusters of age 5-8 Myr which 
exhibited an excess of 
H$\alpha$ emission.  Observational studies suggest that the natal star formation disks of 
early B-type stars can clear on timescales of at least 0.1-0.8 Myr, although it is admittedly not clear if this  
timescale is ``typical'' for all early-type B stars.  The near-IR colors of many of the excess H$\alpha$ emitters 
in these very young clusters are consistent with those 
expected from gaseous classical Be stars and generally 
inconsistent with those observed from dustier B[e] and/or Herbig Ae/Be disk systems.  
The observed wavelength dependence of intrinsic polarization of many H$\alpha$ excess stars in 
clusters of age 5-8 Myr are also consistent with that expected from pure gas disk systems (classical Be 
stars) and exhibit no evidence of secondary contributions from dust scattering, as would be expected 
for composite disk systems (Herbig Ae/Be, B[e] stars).  Based on these factors, we suggest that 
there is compelling evidence supporting the existence of classical Be stars in clusters of age 5-8 Myr.

\subsubsection{Be Stars in the First Half of a B Star's Main Sequence Lifetime: A Comparison to \citet{fab00} and \citet{mar06}}

As outlined and summarized in Section \ref{polconcl}, data presented in this paper support  
the suggestion of WB06 that classical Be stars do develop before the mid-point 
main sequence lifetime of a B star.  Specifically, these data support the presence of a significant 
number of classical Be stars, having crude photometric spectral types 
of B0-B5 (WB06), in clusters spanning a near-continuous range of ages from log(t) of 6.7-8.1, 
i.e. 5-126 Myr.  These results conflict with claims made by \citet{fab00} and \citet{mar06} that the 
Be phenomenon develops in the second-half of a B star's main sequence lifetime.  \citet{mar06} 
do suggest that massive Be stars may briefly appear near the ZAMS in the SMC and 
LMC, before losing their ``Be status'' until the very end of the first part of their main sequence lifetimes.  
We note that this conclusion was essentially 
based on a single cluster population, which limits its robustness.  WB06 identified, and 
the intrinsic polarization 
presented in this paper support, the detection of a large body of Magellanic Cloud Be stars having crude, photometrically-assigned spectral types of B0-B5, in clusters spanning a range of ages from 5-126 Myr.  
No abrupt absence of classical Be stars was observed in any of the younger-type clusters of our photometric 
and/or polarimetric datasets.  Thus, our larger, more complete investigation of Be stars in Magellanic 
Cloud clusters calls into question the suggestion that the Be phenomenon is primarily restricted to the second-half 
of the main-sequence lifetime of B-stars \citep{fab00,mar06} and contradicts the appearance-disappearance-reappearance of the Be phenomenon suggested by \citet{mar06}.    

\subsubsection{Implications of the Presence of Be Stars in Very Young Clusters}  

WB06 first reported the presence of candidate classical Be stars in clusters as young as 5 Myr, 
and noted that if such objects were truly classical Be stars, they would not have spent 
enough time on the main sequence to spin-up to near-critical rotation velocities via the mechanism 
proposed by \citet{mey00} or via mass transfer in a binary system \citep{mcs05}.  The confirmation 
of the classical Be status of many of these stars via the present study can be interpreted as evidence
 that a significant number of classical Be stars emerge onto the 
zero-age-main-sequence at near critical rotation velocities.  An alternate interpretation of our 
results is that these youthful classical Be stars are evidence of the existence of a subset of the Be phenomenon which rotate at significantly sub-critical 
rates, perhaps as low as 0.4-0.6 v$_{crit}$ for early-type B stars \citep{cra05}.  \citet{cra05} suggested that 
only a subset of early-type 
(O7-B2) Be stars might be sub-critical rotators, while later-type objects (B3-A0) should be all near-critical rotators.  
Our photometric survey of these very young clusters assigned crude spectral types to these candidate Be stars of 
B0-B5 (WB06); however, owing to detection biases in our follow-up polarimetric observations, we have in general only been able to confirm that portions of the brightest of these 
candidates (i.e. the earliest spectral sub-types, B0 to $\sim$B3) 
are bona-fide classical Be stars.  Thus our present observations do not rule out the possibility that 
some of our very young bona-fide classical Be stars might belong to the sub-critical rotation population 
predicted by \citet{cra05}.  
  
The observed emergence of the Be phenomenon earlier in the main sequence lifetime than previously thought 
also has important implications regarding the role of magnetic fields in the formation of Be disks \citep{cas03}. 
\citet{mac03} investigated the transport of magnetic flux tubes in 9 M$_{sun}$ stars ($\sim$B2.5-B3) and 
found these structures could rise from the core to the surface within the (at the time) expected beginning 
of the Be phenomenon, the mid-point main sequence lifetime.  The present study has confirmed the presence of bona-fide classical Be stars in clusters as young as 5 Myr having crude, photometrically assigned spectral types of B0 to $\sim$B3.  Although WB06 suggested the presence of an additional population of B4-B5 type candidate Be stars in these clusters, the limited dynamical range probed by the present study was unable to confirm the status of these fainter candidates.  Although the magnetic flux transport 
model of \citet{mac03} had sufficient time to transport flux to the stellar surface for $\sim$B3 stars, this is 
unlikely to be the case for later type stars ($\geq$B4) which have a much thicker radiative envelope.  Hence, the existence of such later-type Be stars in extremely young clusters would likely require an additional 
mechanism to be employed to accelerate the rise times of magnetic flux (Cassinelli, 2006 personal communication).  While challenging, follow-up spectroscopy of these young cluster populations would 
provide one avenue confirm the presence of these purported \citep{wi06a} later-type systems. 

\subsection{Future Work}

While the present study provides significant advances in identifying and understanding the biases present in
earlier 2-CD studies of cluster populations, additional work is clearly required.  Our polarimetric survey 
was not sensitive to pole-on or near-pole-on systems, hence determining which of our 
``type-2'' objects are near-pole-on classical Be stars and which are astrophysical objects of a 
fundamentally different nature is a high priority.  We suggest that followup moderate resolution optical 
spectroscopic or infrared photometric observations would provide reasonable diagnostics to resolve this
bias, and enable more quantitative determinations of the bona-fide classical Be content of clusters as a function of
age and/or metallicity.  Our data suggest that at least 25\% of photometrically identified candidate Be 
stars appear unlikely to be true classical Be stars; however, we are unable to place firm constraints on the
true astrophysical nature of such objects.  We suggest that followup infrared photometric and/or spectroscopic
observations would be useful to further constrain the evolutionary nature of these objects.  Furthermore, such 
observations would also be useful to ascertain the true nature of candidate Be stars which were too faint to be
reliably probed by our polarimetric observations.

Our analysis of the intrinsic polarization properties of LMC and SMC classical Be stars suggests that the
fundamental disk properties of classical Be stars may depend on metallicity.  Modeling followup moderate 
resolution spectropolarimetric observations of LMC/SMC classical Be stars is a clear avenue one could use to
further investigate these results, and we note that the advent of accurate linear spectropolarimeters on large 
aperture telescopes (i.e. the Focal Reducer Spectrograph (FORS) at the Very Large Telescope (VLT) or the Robert Stobie Spectrograph at the South African Large Telescope (SALT)) make obtaining such observations feasible.

\section{Summary}

We have used imaging polarimetric observations of six LMC and six SMC clusters to investigate the evolutionary
status of B-type stars identified as excess H$\alpha$ emitters via optical 2-color diagram photometric 
techniques.  We characterized the interstellar polarization components of these data in a systematic manner, 
allowing us to isolate the intrinsic polarization properties of our data, hence constrain the dominant polarigenic 
agent in each system.  We found: \begin{enumerate}
\item The interstellar polarization associated with NGC 330 was characterized by $\lambda_{max}$ $\sim$4500\AA, 
suggesting the presence of a small dust grain population; 

\item The ISP of NGC 2100 exhibited clear evidence of a complex morphology, indicating the presence of a 
non-uniform magnetic field.  We offer a detailed discussion of this B field and its potential origin in \citet{wi06b}.  

\item The UBVRI wavelength dependence of \textit{intrinsic} polarization of many candidate Be stars in 
clusters of age 5-8 Myr exhibit either polarization Balmer jumps and/or electron scattering signatures, which 
is the expected diagnostic of free-free scattering from a gaseous disk.  No evidence of a secondary polarigenic 
agent, i.e. dust scattering, is observed in these system.  Moreover, the 2MASS near-IR colors of many of these 
systems are most consistent with the expected colors of pure gas disks, and inconsistent with the observed colors 
of dustier Magellanic Cloud disk systems (Herbig Ae/Be, B[e]).  We conclude these data confirm the initial 
suggestion of \citet{wi06a} that classical Be stars are present in clusters of age 5-8 Myr, contradicting claims 
that the Be phenomenon only develops in the second half of a B star's main sequence lifetime 
\citep{fab00,mar06}, i.e. after 10 Myr.  

\item We interpret the observed presence of classical Be stars in clusters of age 5-8 Myr as evidence that 
a significant population of early-type B stars must emerge onto the ZAMS rotating at near critical velocities.  
We note however that these results do not exclude the possibility that we are observing the hypothesized subset of 
classical Be stars which rotate at sub-critical velocities \citep{cra05}.  

\item Comparing the polarimetric properties 
of our dataset to a similar survey of Galactic classical Be stars, we find the prevalence of polarimetric 
Balmer jump signatures decreases with metallicity.  We speculate that these results might indicate that 
either it is more difficult to form large disk systems in low metallicity environments, or that the average 
disk temperature is higher in these low metallicity environments.  

\item We find evidence that at least 25\% of
photometrically identified candidate Be stars do not exhibit polarimetric signatures consistent with those 
expected from classical Be stars.  These data strongly suggest that caution must be exercised when 
attempting to correlate the onset of the Be phenomenon with evolutionary age and/or metallicity based solely 
on statistics derived from simple 2-color diagram photometry.

\end{enumerate}

\acknowledgments

We thank Joe Cassinelli, Doug Gies, and Jennifer Hoffman for discussions which 
enhanced this 
paper.  The efforts of Brian Babler, Ken Nordsieck, and the PBO observing
crew for supplying the HPOL data presented here are greatly appreciated.  We thank the NOAO TAC 
for allocating observing time for this project and JPW thanks NOAO for supporting his travel to 
CTIO.  Portions of this work has been supported in part by NASA NPP NNH06CC03B (JPW), NASA 
GRSP NGT5-50469 (JPW), NASA LTSA NAG5-8054 (KSB), NSF AST-0307686 (JEB), and FAPESP 
02/12880-0 (AP) grants.  AMM acknowledges travel support by FAPESP; 
he is also partially supported by CNPq.  Polarimetry at the University of S\~ao Paulo (USP) is 
supported by FAPESP.  This research has made use of the SIMBAD database operated 
at CDS, Strasbourg, France, and the NASA ADS system.

\newpage
\begin{table}
\begin{center}
\scriptsize
\caption{Summary of our observations \label{polexpose}}
\begin{tabular}{lccccc}
\tableline
Cluster & Location & Filter  & Age & Date & Exposure Time \\ 
\tableline
Bruck 60 & SMC & U & o & 2002 Oct 19 & 1200 \\
\nodata & \nodata & B & \nodata & 2002 Oct 21 & 720 \\
\nodata & \nodata & V & \nodata & 2002 Oct 19 & 300 \\
\nodata & \nodata & R & \nodata & 2002 Oct 19 & 300 \\
\nodata & \nodata & I & \nodata & 2002 Oct 21 & 300 \\
NGC 330   & SMC & U & y & 2001 Nov 22 & 1200 \\
\nodata & \nodata & B & \nodata & 2001 Nov 23 & 420 \\
\nodata & \nodata & V & \nodata & 2001 Nov 25 & 300 \\
\nodata & \nodata & R & \nodata & 2001 Nov 25 & 240 \\
\nodata & \nodata & I & \nodata & 2001 Nov 24 & 240 \\
NGC346    & SMC & U & vy & 2001 Nov 21 & 720 \\
\nodata & \nodata & B & \nodata & 2001 Nov 22 & 240 \\
\nodata  & \nodata & V & \nodata & 2001 Nov 21 & 180 \\
\nodata & \nodata & R & \nodata & 2001 Nov 24 & 180 \\
\nodata & \nodata & I & \nodata & 2001 Nov 23 & 180 \\
NGC 371 & SMC & U & vy & 2002 Oct 18 & 1200 \\
\nodata & \nodata & B & \nodata & 2002 Oct 18 & 600 \\
\nodata & \nodata & V & \nodata & 2002 Oct 17 & 420 \\
\nodata & \nodata & R & \nodata & 2002 Oct 17 & 300 \\
\nodata & \nodata & I & \nodata & 2002 Oct 18 & 420 \\
NGC 456 & SMC & U & y & 2002 Oct 24 & 1200 \\
\nodata & \nodata & B & \nodata & 2002 Oct 26 & 720 \\
\nodata & \nodata & V & \nodata & 2002 Oct 24 & 420 \\
\nodata & \nodata & R & \nodata & 2002 Oct 26 & 300 \\
\nodata & \nodata & I & \nodata & 2002 Oct 25 & 300 \\
NGC 458 & SMC & U & o & 2002 Oct 27 & 1200 \\
\nodata & \nodata & B & \nodata & 2002 Oct 27 & 720 \\
\nodata & \nodata & V & \nodata & 2002 Oct 26 & 300 \\
\nodata & \nodata & I & \nodata & 2002 Oct 27 & 300 \\
LH 72 & LMC & U & vy & 2002 Oct 26 & 1200 \\
\nodata & \nodata & B & \nodata & 2002 Oct 25 & 720 \\
\nodata & \nodata & V & \nodata & 2002 Oct 25 & 420 \\
\nodata & \nodata & R & \nodata & 2002 Oct 27 & 360 \\
\nodata & \nodata & I & \nodata & 2002 Oct 26 & 300 \\
NGC 1818 & LMC & U & y & 2001 Nov 21 & 900 \\
\nodata & \nodata & B & \nodata & 2001 Nov 22 & 240 \\
\nodata & \nodata & V & \nodata & 2001 Nov 21 & 240 \\
\nodata & \nodata & R & \nodata & 2001 Nov 22 & 180 \\
\nodata & \nodata & I & \nodata & 2001 Nov 22 & 300 \\
NGC 1858 & LMC & U & vy & 2002 Oct 19 & 1200 \\
\nodata & \nodata & B & \nodata & 2002 Oct 24 & 720 \\
\nodata & \nodata & V & \nodata & 2002 Oct 18 & 420 \\
\nodata & \nodata & R & \nodata & 2002 Oct 21 & 300 \\
\nodata & \nodata & I & \nodata & 2002 Oct 24 & 300 \\
NGC 1948 & LMC & U & y & 2001 Nov 27 & 1200 \\
\nodata & \nodata & B & \nodata & 2001 Nov 26 & 360 \\
\nodata & \nodata & V & \nodata & 2001 Nov 25 & 180 \\
\nodata & \nodata & R & \nodata & 2001 Nov 26 & 240 \\
\nodata & \nodata & I & \nodata & 2001 Nov 26 & 300 \\
NGC 2004 & LMC & U & y & 2001 Nov 25 & 900 \\
\nodata & \nodata & B & \nodata & 2001 Nov 22 & 180 \\
\nodata & \nodata & V & \nodata & 2001 Nov 27 & 180 \\
\nodata & \nodata & I & \nodata & 2001 Nov 22 & 420 \\
NGC 2100 & LMC & U & y & 2001 Nov 24 & 1200 \\
\nodata & \nodata & B & \nodata & 2001 Nov 23 & 240 \\
\nodata & \nodata & V & \nodata & 2001 Nov 23 & 180 \\
\nodata & \nodata & R & \nodata & 2001 Nov 24 & 180 \\
\nodata & \nodata & I & \nodata & 2001 Nov 23 & 180 \\
\tableline
\tablecomments{We summarize some of the properties of the data presented in this paper.  The exposure 
times listed are the integration time used at each of eight
wave-plate position.  The age labels in column 4, originally defined and discussed 
in detail in \citet{wi06a}, correspond to very young (vy) 5-8 Myr; young (y) 10-25 Myr; 
and old (o) 32-158 Myr.}
\end{tabular}
\end{center}
\end{table}

\newpage
\begin{table}
\begin{center}
\scriptsize
\caption{Average interstellar polarization \label{isptable}}
\begin{tabular}{lcccccccccc}
\tableline\tableline
 Cluster & \% P$_{u}$ & \% P$_{b}$ & \% P$_{v}$ & \% P$_{r}$ &
\% P$_{i}$ & PA & P$_{max}$ & $\lambda_{max}$ & dPA & K \\
\tableline
Bruck 60 & 0.46$\pm$0.18\% & 0.51$\pm$0.30\% & 0.55$\pm$0.23\% & 0.54$\pm$0.19\% & 0.48$\pm$0.19\% & 130 & 0.55\% & 
 5500 & 0 & 0.923 \\
LH 72   & 0.34$\pm$0.14\% & 0.37$\pm$0.12\% & 0.40$\pm$0.14\% & 0.39$\pm$0.12\% & 0.35$\pm$0.14\% & 28  & 0.40\% &
 5500 & 0 & 0.923 \\
NGC 330 & 0.51$\pm$0.17\% & 0.53$\pm$0.16\% & 0.52$\pm$0.14\% & 0.48$\pm$0.17\% & 0.41$\pm$0.17\% & 126 & 0.53\% &
 4500 & 0 & 0.737 \\
NGC 346 & 0.31$\pm$0.17\% & 0.35$\pm$0.18\% & 0.37$\pm$0.17\% & 0.36$\pm$0.15\% & 0.32$\pm$0.19\% & 125 & 0.37\% &
 5500 & 0 & 0.923 \\
NGC 371 & 0.38$\pm$0.15\% & 0.42$\pm$0.16\% & 0.45$\pm$0.15\% & 0.44$\pm$0.15\% & 0.39$\pm$0.15\% & 118 & 0.45\% &
 5500 & 0 & 0.923 \\
NGC 456 & 0.57$\pm$0.21\% & 0.63$\pm$0.24\% & 0.67$\pm$0.25\% & 0.66$\pm$0.29\% & 0.58\% & 147 & 0.67\% &
 5500 & -8 & 0.923 \\
NGC 458 & 0.33$\pm$0.34\% & 0.36$\pm$0.23\% & 0.39$\pm$0.22\% & \nodata & 0.34$\pm$0.15\% & 120 & 0.39\% &
 5500 & 0 & 0.923 \\
\tableline
NGC 1818 & 0.52$\pm$0.17\% & 0.58$\pm$0.16\% & 0.62$\pm$0.15\% & 0.61$\pm$0.17\% & 0.54$\pm$0.14\% & 39 & 0.62\% &
 5500 & 0 & 0.923 \\
NGC 1858 & 0.32$\pm$0.21\% & 0.36$\pm$0.15\% & 0.38$\pm$0.16\% & 0.37$\pm$0.22\% & 0.33$\pm$0.17\% & 45 & 0.38\% &
 5500 & 0 & 0.923 \\
NGC 1948 & 0.57$\pm$0.19\% & 0.64$\pm$0.15\% & 0.68$\pm$0.18\% & 0.67$\pm$0.17\% & 0.59$\pm$0.17\% & 35 & 0.68\% &
 5500 & 0 & 0.923 \\
NGC 2004 & 0.32$\pm$0.12\% & 0.36$\pm$0.13\% & 0.38$\pm$0.15\% & \nodata & 0.33$\pm$0.16\% & 26 & 0.38\% &
 5500 & 0 & 0.923 \\
\tableline\tableline
\tablecomments{The Serkowski parameters PA, P$_{max}$, $\lambda_{max}$, 
dPA, and K which best described each of our clusters are tabulated, along with the final ISP values in the 
U, B, V, R, and I filters.  The horizontal line separates SMC clusters (above) from LMC clusters (below).}
\end{tabular}
\end{center}
\end{table}

\newpage
\begin{table}
\begin{center}
\scriptsize
\caption{Foreground ISP Corrections \label{foreground}}
\begin{tabular}{lccc}
\tableline\tableline
Schmidt Region & Polarization & Position Angle & Clusters \\
\tableline

Region I & 0.37 $\pm$ 0.15\% & 111 & Bruck 60 \\ 
Region II & 0.27$\pm$0.15\% & 123 & NGC 330, 346, 371 \\
Region III & 0.19$\pm$0.21\% & 110 & NGC 456, 458 \\
Region VI & 0.64$\pm$0.19\% & 30 & LH 72 \\
Region VII & 0.32$\pm$0.13\% & 29 & NGC 1818,1858,1948 \\
Region IX & 0.40$\pm$0.13\% & 20 & NGC 2004 \\
\tableline\tableline
\tablecomments{We used the Galactic polarization maps of \citet{sch76} as templates to identify and 
remove the foreground Galactic interstellar polarization contribution to each of our Magellanic Cloud 
lines of sight.  Columns two and three represent the corrections applied to each of our lines of sight, 
listed in column four.}
\end{tabular}
\end{center}
\end{table}

\newpage
\begin{table}
\begin{center}
\scriptsize
\caption{ISP Intrinsic to the Magellanic Clouds \label{ispintrin}}
\begin{tabular}{lcccccccccc}
\tableline\tableline
Cluster & P$_{u}$ & PA$_{u}$ & P$_{b}$ & PA$_{b}$ &
P$_{v}$ & PA$_{v}$ & P$_{r}$ & PA$_{r}$ & P$_{i}$ & PA$_{i}$ \\
\tableline

Bruck 60 & 0.43$\pm$0.23\% & 132$\pm$15 & 0.45$\pm$0.34\% & 133$\pm$22 & 
0.34$\pm$0.27\% & 151$\pm$23 & 0.41$\pm$0.24\% & 138$\pm$17 & 0.33$\pm$0.24\% & 144$\pm$21 \\
LH 72 & 0.27$\pm$0.24\% & 28$\pm$26 & 0.24$\pm$0.22\% & 27$\pm$26 & 0.24$\pm$0.24\% & 
123$\pm$29 & 0.07$\pm$0.22\% & 18$\pm$90 & 0.07$\pm$0.24\% &  130$\pm$98 \\
NGC 330 & 0.48$\pm$0.23\% & 126$\pm$14 & 0.48$\pm$0.22\% & 126$\pm$13 & 0.25$\pm$0.21 & 
129$\pm$24 & 0.34$\pm$0.23\% & 127$\pm$19 & 0.24$\pm$0.23\% &  128$\pm$28 \\
NGC 346 & 0.28$\pm$0.23\% & 125$\pm$24 & 0.30$\pm$0.23\% & 125$\pm$22 & 
0.10$\pm$0.23\% & 130$\pm$66 & 0.22$\pm$0.21\% & 126$\pm$27 & 0.15$\pm$0.24\% & 127$\pm$46 \\
NGC 371 & 0.35$\pm$0.21\% & 118$\pm$17 & 0.37$\pm$0.22\% & 117$\pm$17 & 0.19$\pm$0.21\% & 
111$\pm$32 & 0.31$\pm$0.21\% & 116$\pm$19 & 0.22$\pm$0.21\% & 114$\pm$27 \\
NGC 456 & 0.56$\pm$0.30\% & 148$\pm$15 & 0.62$\pm$0.32\% & 149$\pm$15 & 0.64$\pm$0.33\% & 
155$\pm$15 & 0.64$\pm$0.36\% & 151$\pm$16 & 0.56\% & 153.2 \\
NGC 458 & 0.31$\pm$0.40\% & 121$\pm$37 & 0.32$\pm$0.31\% & 121$\pm$28 & 0.22$\pm$0.30\% &  
129$\pm$39 & 0.23$\pm$0.26\% & 125$\pm$32 & \nodata & \nodata \\
\tableline
NGC 1818 & 0.49$\pm$0.21\% &  40$\pm$12 & 0.52$\pm$0.21\% & 40$\pm$12 & 0.34$\pm$0.20\% &  
49$\pm$17 & 0.46$\pm$0.21\% & 43$\pm$13 & 0.35$\pm$0.19\% & 45$\pm$16 \\
NGC 1858 & 0.29$\pm$0.25\% & 47$\pm$25 & 0.31$\pm$0.20\% & 48$\pm$19 & 0.20$\pm$0.21\% & 
74$\pm$30 & 0.25$\pm$0.26\% & 55$\pm$30 & 0.19$\pm$0.21\% & 63$\pm$32 \\
NGC 1948 & 0.54$\pm$0.23\% &  35$\pm$12 & 0.58$\pm$0.20\% & 36$\pm$10 & 0.37$\pm$0.22\% & 
40$\pm$17 & 0.51$\pm$0.21\% & 37$\pm$12 & 0.39$\pm$0.21\% & 38$\pm$15 \\
NGC 2004 & 0.28$\pm$0.18\% & 27$\pm$18 & 0.28$\pm$0.18\% & 28$\pm$18 & 0.08$\pm$0.20\% & 
75$\pm$72 & 0.09$\pm$0.21\% & 44$\pm$67 & \nodata & \nodata \\
\tableline\tableline
\tablecomments{We corrected the total measured ISP components of each of our clusters for the effects of 
foreground Galactic interstellar polarization, revealing the ISP components which arose from the dichroic 
absorption of light by dust grains located
within the Magellanic Clouds. The horizontal line separates SMC clusters (above) from LMC clusters (below).}
\end{tabular}
\end{center}
\end{table}

\newpage
\begin{table}
\begin{center}
\scriptsize
\caption{Total interstellar polarization toward NGC 2100 \label{isptable2}}
\begin{tabular}{lcccccccccc}
\tableline\tableline
 Cluster & \% P$_{u}$ & \% P$_{b}$ & \% P$_{v}$ & \% P$_{r}$ &
\% P$_{i}$ & PA & P$_{max}$ & $\lambda_{max}$ & dPA & K \\
\tableline
NGC 2100 area 1 & 1.32\% & 1.41\% & 1.44\% & 1.37\% & 1.20\% & 139 & 1.45\% &
 5000 & 0 & 0.83 \\
NGC 2100 area 2 & 1.26\% & 1.35\% & 1.37\% & 1.31\% & 1.14\% & 100 & 1.38\% &
5000 & 0 & 0.83 \\
NGC 2100 area 3 & 1.64\% & 1.69\% & 1.65\% & 1.54\% & 1.32\% & 78 & 1.70\% & 
4500 & 0 & 0.737 \\
\tableline\tableline
\tablecomments{A summary of the total interstellar polarization values toward NGC 2100 
for the 3 distinct spatial
regions identified within the field of view of our NGC 2100 dataset.}
\end{tabular}
\end{center}
\end{table}

\newpage
\begin{table}
\begin{center}
\scriptsize
\caption{Total polarization of candidate Be stars \label{totpol}}
\begin{tabular}{lcccccc}
\tableline
Candidate Be Star & Filter & \%P & PA & \% Q & \% U & \% Err \\ 
\tableline
Bruck 60:WBBe 1 & u &  1.75 & 107.3 & -1.44 & -0.99 &  0.61 \\
   & b &  \nodata &  \nodata &  \nodata &  \nodata &  \nodata \\
   & v &  1.33 & 113.7 & -0.90 & -0.98 &  0.30 \\
   & r &  1.33 & 117.6 & -0.76 & -1.09 &  0.35 \\
   & i &  \nodata & \nodata &  \nodata &  \nodata & \nodata \\ 
\tableline
\tablecomments{The total polarization (i.e. interstellar plus intrinsic components) of
one candidate Be star investigated in this study is tabulated.  A table presenting results for 
our entire dataset is available in the electronic version of this Journal.}
\end{tabular}
\end{center}
\end{table}

\newpage
\begin{table}
\begin{center}
\scriptsize
\caption{Intrinsic polarization of candidate Be stars \label{inpol}}
\begin{tabular}{lcccccc}
\tableline
Candidate Be Star & Filter & \%P & PA & \% Q & \% U & \% Err \\ 
\tableline
Bruck 60:WBBe 1 & u &  1.46 & 100.9 & -1.36 & -0.54 &  0.61 \\
   & b &  \nodata &  \nodata &  \nodata &  \nodata &  \nodata \\
   & v &  0.91 & 104.4 & -0.80 & -0.44 &  0.30 \\
   & r &  0.88 & 110.0 & -0.67 & -0.56 &  0.35 \\
   & i &  \nodata & \nodata & \nodata & \nodata & \nodata \\ 
\tableline
\tablecomments{The intrinsic polarization of
one candidate Be star investigated in this study is tabulated.  A table presenting results for 
our entire dataset is available in the electronic version of this Journal.}
\end{tabular}
\end{center}
\end{table}

\newpage
\begin{table}
\begin{center}
\scriptsize
\caption{Classifying the intrinsic polarization of individual candidate Be stars \label{starclass}}
\begin{tabular}{lcccccc}
\tableline
Name & Pol Balmer Jump? & PA & Classification & Comments \\ 
\tableline
Bruck 60:WBBe 1  &    N  & F2 & 2  & moderate ES \\
\tableline
\tablecomments{The classification of the intrinsic polarization signature of one of the candidate Be 
stars investigated in this study is presented.  Column 2 indicates whether a polarization Balmer 
jump was present in the data; the letters in column 3 indicate whether the polarization position 
angle was flat (F), linearly sloped (S), or randomly variable (V) while the numbers indicate the 
$\sigma$ level of such characterizations; column 4 indicates the likelihood each object is a bona-fide 
classical Be star based on the 4 point classification scheme described in the text; and any 
relevant comments are listed in column 5.  Note that a table presenting results for 
our entire dataset is available in the electronic version of this Journal.}
\end{tabular}
\end{center}
\end{table}

\newpage
\begin{table}
\begin{center}
\scriptsize
\caption{Intrinsic polarization summary \label{polsummary}}
\begin{tabular}{lccccc}
\tableline\tableline
Cluster & Detection Rate & \# Pol. BJ & \# ES & Not Inconsistent  & Unlikely \\
\nodata & \nodata & (type-1) & (type-2) & (type-2) & (type-3,4) \\
\tableline
Bruck 60 & 18/26 (69\%) & 1/18 (6\%) & 8/18 (44\%) & 11/18 (61\%) & 6/18 (33\%) \\
NGC 330  & 41/76 (54\%) & 0/41 (0\%) & 9/41 (22\%) & 24/41 (59\%) & 17/41 (41\%) \\
NGC 346  & 33/48 (69\%) & 8/33 (24\%)& 7/33 (21\%) & 22/33 (67\%) & 3/33 (9\%) \\
NGC 371  & 73/129 (57\%) & 10/73 (14\%) & 6/73 (8\%) & 49/73 (67\%) & 14/73 (19\%) \\
NGC 456  & 14/23 (61\%) & 0/14 (0\%) & 1/14 (7\%) & 9/14 (64\%) & 5/14 (36\%) \\
NGC 458  & 10/30 (33\%) & 0/10 (0\%) & 0/10 (0\%) & 6/10 (60\%) & 4/10 (40\%) \\
\tableline
LH 72    & 34/50 (67\%) & 1/34 (3\%) & 5/34 (15\%) & 22/34 (65\%) & 11/34 (32\%) \\
NGC 1818 & 18/40 (45\%) & 4/18 (22\%) & 5/18 (28\%) & 12/18 (67\%) & 2/18 (11\%) \\
NGC 1858 & 27/39 (69\%) & 3/27 (11\%) & 2/27 (7\%) & 13/27 (48\%) & 11/27 (41\%) \\
NGC 1948 & 22/27 (81\%) & 6/22 (27\%) & 6/22 (27\%) & 12/22 (55\%) & 4/22 (18\%) \\
NGC 2004 & 43/67 (64\%) & 9/43 (21\%) & 10/43 (23\%) & 28/43 (65\%) & 6/43 (14\%) \\
NGC 2100 & 35/61 (57\%) & 8/35 (23\%) & 16/35 (46\%) & 20/35 (57\%) & 7/35 (20\%) \\
\tableline\tableline
\tablecomments{We summarize the statistics of our classification of the intrinsic polarization properties 
of our dataset.  The detection rate column describes
the number of candidates for which we were able to extract polarimetric
information.  The third column denotes the fraction of candidates which
display polarimetric Balmer jumps (BJ): we consider these objects to be bona-fide classical Be stars.  The
fourth column describes the fraction of candidates which exhibit electron scattering (ES) signatures, 
i.e. moderate amounts of wavelength-independent
polarization at a wavelength independent position angle.  The fifth column
describes candidates that, to within 3 $\sigma$, can not be ruled out as
possible Be stars.  The sixth column 
describes candidate Be stars whose intrinsic polarization does not agree,
to within 3 $\sigma$, with that expected from classical Be stars.  The horizontal line separates SMC clusters (above) from LMC clusters (below).}
\end{tabular}
\end{center}
\end{table}

\newpage
\begin{table}
\begin{center}
\scriptsize
\caption{Average intrinsic polarization properties \label{polsummary2}}
\begin{tabular}{lcccccc}
\tableline\tableline
Cluster & Mean BJ & Median BJ & Mean ES & Median ES & Mean Unlikely & Median Unlikely \\
\tableline
Very Young & 13\% & 13\% & 13\% & 12\% & 25\% & 26\% \\
Young & 13\% & 21\% & 22\% & 23\% & 26\% & 20\% \\
\tableline
SMC & 7\% & 3\% & 17\% & 15\% & 30\% & 35\% \\
LMC & 18\% & 22\% & 24\% & 25\% & 23\% & 19\% \\
\tableline
ALL SMC+LMC & 13\% & 13\% & 21\% & 22\% & 26\% & 26\% \\
\tableline
\tablecomments{Statistical averages of data listed in Table \ref{polsummary} are presented.  Clusters 
were grouped according to age (very young and young) as well as metallicity (SMC and LMC).  The 
final entry labeled \textit{ALL} denotes a global average of all SMC and LMC cluster data regardless of age or metallicity.}
\end{tabular}
\end{center}
\end{table}

\newpage
\begin{table}
\begin{center}
\caption{Average intrinsic polarization properties with ``contaminants'' removed \label{polsum3}}
\begin{tabular}{lcccc}
\tableline\tableline
Cluster & Mean BJ & Median BJ & Mean ES & Median ES \\
\tableline
Very Young & 17\% & 18\% & 17\% & 18\% \\
Young & 16\% & 24\% & 28\% & 31\% \\
\tableline
SMC & 7\% & 4\% & 25\% & 17\% \\
LMC & 22\% & 25\% & 31\% & 29\% \\
\tableline
ALL LMC+SMC & 16\% & 18\% & 28\% & 25\% \\
Milky Way Galaxy & 42-49\% & \nodata & \nodata & \nodata \\
\tableline
\tablecomments{Objects deemed unlikely to be classical Be stars in Table \ref{polsummary2} were 
removed from consideration, and the statistical averages of data listed in Table \ref{polsummary} 
were recalculated.  The ``ALL SMC+LMC'' entry denotes a global average of all SMC and LMC cluster 
data regardless of age or metallicity.  The Milky Way data averages were derived from 
analysis of the HPOL polarization catalog, as discussed in Section 5.1.}
\end{tabular}
\end{center}
\end{table}

\newpage
\begin{table}
\begin{center}
\scriptsize
\caption{Frequency of Contaminants in Be Populations Identified via 2-CDs \label{photcontam}}
\begin{tabular}{lcc}
\tableline\tableline
Cluster & Analysis of 2-CD data & Analysis of Polarimetry \\
\tableline
Bruck 60 & 14\% & 33\% \\
NGC 371  & 11\% & 19\% \\
NGC 456  & 10\% & 36\% \\
NGC 458  & 38\% & 40\% \\
LH 72    & 25\% & 32\% \\
NGC 1858 & 42\% & 41\% \\
\tableline\tableline
\tablecomments{The possible rate of contamination of classical Be star detections extracted from 
photometric 2-color diagrams.  As discussed in Section 5.2, the second column represents the 
frequency of red-type excess H$\alpha$ emitters, which we suggest might be an appropriate proxy 
of the frequency of B-type excess H$\alpha$ emitters which are not classical Be stars.  The third 
column is a reproduction of column 6 in Table \ref{polsummary} and illustrates the frequency of 
2-CD contaminants as derived from polarimetric observations of candidate Be stars.}
\end{tabular}
\end{center}
\end{table}

\newpage
\begin{table}
\begin{center}
\scriptsize
\caption{IR Colors of Candidate Be Stars \label{ircolor}}
\begin{tabular}{lccccc}
\tableline\tableline
Star & Classification & (J-H) & (H-K) & (J-H)$_{err}$ & (H-K)$_{err}$ \\
\tableline
Bruck60:WBBe 5    & Type2-4,other & 0.09 & 0.04 & 0.14 & 0.24 \\
Bruck60:WBBe 20 &  Type2-4,other & 0.29 & 0.00 & 0.13 & 0.20 \\
LH72:WBBe 5     &  Type1:Bal-J & 0.07 & 0.21 & 0.10 & 0.15 \\
LH72:WBBe 15   &   Type2:ES  &  0.54 & 0.24 & 0.14 & 0.17 \\

\tableline\tableline
\tablecomments{Near-IR colors are tabulated for all candidate Be stars in clusters 
investigated in this paper which were detected in all three 2MASS photometric bands.  The 
quoted color errors were obtained by adding individual filter errors in quadrature.  Column 2 
designates the polarimetric designation, if any, assigned to each target within this paper; ``Type-1:Bal-J'' 
designates Type-1 sources which we assert are most likely to be classical Be stars, ``Type-2:ES'' 
designates Type-2 sources which exhibit electron scattering intrinsic polarimetric signatures and are 
thus likely classical Be stars, and ``Type-2-4,other'' designates all other sources, regardless of whether 
they were detected via our polarimetric survey or not.  Note that a table presenting results for 
our entire dataset is available in the electronic version of this Journal.}
\end{tabular}
\end{center}
\end{table}

\newpage

\newpage

\newpage

\newpage
\clearpage

\newpage

\newpage

\newpage

\newpage

\newpage

\newpage

\newpage

\newpage

\newpage
\clearpage

\newpage

\newpage

\newpage

\newpage

\end{document}